\documentclass[11pt]{article}

\usepackage{aas_macros,amsmath,amssymb,comment,cite,esint,graphicx,mathtools}
\usepackage[margin=.8in,letterpaper]{geometry}
\usepackage[colorlinks=true]{hyperref}
\usepackage[affil-it]{authblk}
\usepackage{subcaption}
\usepackage[utf8]{inputenc}
\usepackage{mathrsfs}
\usepackage{appendix}
\usepackage{amssymb}
\usepackage{float}
\usepackage{color}
\usepackage{cite}
\usepackage{hyperref}
\hypersetup{pageanchor=false}
\usepackage{indentfirst}
\usepackage{url}
\usepackage{float}
\usepackage{caption}
\usepackage[numbers,square,comma,sort&compress,merge]{natbib}
\usepackage{esint}
\usepackage{overpic}
\usepackage{graphicx}
\usepackage{epsf,amsmath,bbold,amsfonts,stmaryrd}
\usepackage{textcomp}
\usepackage{ulem}
\usepackage{tikz}

\numberwithin{equation}{section}
\let\originalleft\left
\let\originalright\right
\renewcommand{\left}{\mathopen{}\mathclose\bgroup\originalleft}
\renewcommand{\right}{\aftergroup\egroup\originalright}
\mathcode`\*="8000
{\catcode`\*=\active\gdef*{\mathclose{}\,\mathopen{}}}

\newcommand{\be}{\begin{equation}}
\newcommand{\ee}{\end{equation}}
\newcommand{\bea}{\setlength\arraycolsep{2pt} \begin{eqnarray}}
\newcommand{\eea}{\end{eqnarray}}
\newcommand{\nn}{\nonumber}

\newcommand{\mP}{{\mathcal P}}

\newcommand{\md}{\mathrm{d}}

\def\a{\alpha}
\def\b{\beta}

\def\d{\delta}

\def\f{\frac}

\def\m{\mu} 
\def\n{\nu} 
\def\nn{\nonumber}
\def\pl{\partial}
\def\p{\phi}

\def\t{\theta}

\def\k{\kappa}

\def\be{\begin{equation}}
\def\ee{\end{equation}}
\def\bag{\begin{aligned}}
\def\eag{\end{aligned}}
\def\bea{\begin{eqnarray}}
\def\eea{\end{eqnarray}}
\def\ba{\begin{array}}
\def\ea{\end{array}}

\def\bc{\begin{center}}
\def\ec{\end{center}}

\begin{document}

\title{\textbf {Hybrid Black Hole-- and Disk-Driven Jets: 
\\Steady Axisymmetric Ideal MHD Modeling}}

\author{Yu Song$^{1}$\thanks{songyuphy@gmail.com}~, Yehui Hou$^{2}$\thanks{corresponding author: yehuihou@sjtu.edu.cn}~,  
Lei Huang$^{3, 4}$\thanks{muduri@shao.ac.cn}~, Bin Chen$^{5,1,6}$\thanks{corresponding author: chenbin1@nbu.edu.cn}~}
\date{}
	
\maketitle
\vspace{-10mm}

\begin{center}
{\it
$^1$ School of Physics, Peking University, No.5 Yiheyuan Rd, Beijing
100871, P.R. China\\\vspace{4mm}

$^2$ Tsung-Dao Lee Institute, Shanghai Jiao-Tong University, Shanghai, 201210, P. R. China\\\vspace{4mm}

$^3$ Shanghai Astronomical Observatory, Chinese Academy of Sciences, Shanghai, 200030, P. R. China \\\vspace{4mm}

$^4$ State Key Laboratory of Radio Astronomy and Technology, A20 Datun Road, Chaoyang District, Beijing, 100101, P. R. China \\\vspace{4mm}

$^5$ Institute of Fundamental Physics and Quantum Technology,\\ \& School of Physical Science and Technology, Ningbo University, Ningbo, Zhejiang 315211, China\\\vspace{4mm}

$^6$ Center for High Energy Physics, Peking University,
No.5 Yiheyuan Rd, Beijing 100871, P. R. China\\\vspace{4mm}

}
\end{center}

\vspace{8mm}

\begin{abstract}

Improved observations of relativistic jets have underscored the need for tractable theoretical models. Here we develop a semi-analytical hybrid jet model that combines black hole-driven and disk-driven components within steady, axisymmetric, ideal general relativistic magnetohydrodynamics framework. We derive a criterion for the launching sites of cold outflows, introducing a new constraint on the magnetic field configuration threading a thin disk. Using the Bernoulli equation and critical-point analysis under Michel's minimal-energy ansatz, we obtain flow solutions along different classes of field lines. The hybrid model shows that discontinuities in field-line angular velocity produce pronounced velocity shear and density jumps at the interface between the two jet components, together with localized enhancements that may account for observed limb brightening.

\end{abstract}

\maketitle

\newpage
\baselineskip 18pt

\tableofcontents

\section{Introduction}

Relativistic jets from spinning black holes rank among the most energetic phenomena in astrophysics \cite{frank2002accretion, Piran:2004ba, meier2012black, Kumar:2014upa}. These bipolar outflows, extending over several orders of magnitude in length along the spin axis, have been observed across the electromagnetic spectrum—from radio to gamma rays—in systems such as active galactic nuclei (AGNs), X-ray binaries, and gamma-ray bursts \cite{Blandford:1979za, Pudritz:2012xj}, and exhibits typical shape transition \cite{asada2012structure, Kovalev:2019cue}. A prominent example is the jet from M87*, which exhibits relativistic motion on kiloparsec scales \cite{baade1979identification, owen1989high, asada2012structure}. 
Recent high-resolution observations have resolved the jet base down to horizon scales, revealing edge brightening and possible precession signatures \cite{Lu:2023bbn, Cui:2023uyb}.
Next-generation Very Long Baseline Interferometry facilities, including the next-generation Event Horizon Telescope (ngEHT) and the Black Hole Explorer (BHEX), are expected to resolve even finer jet structures \cite{Ayzenberg:2023hfw,Johnson:2023ynn,Johnson:2024ttr}.

\begin{figure}[ht!]
	\centering
	\includegraphics[width=3.2in]{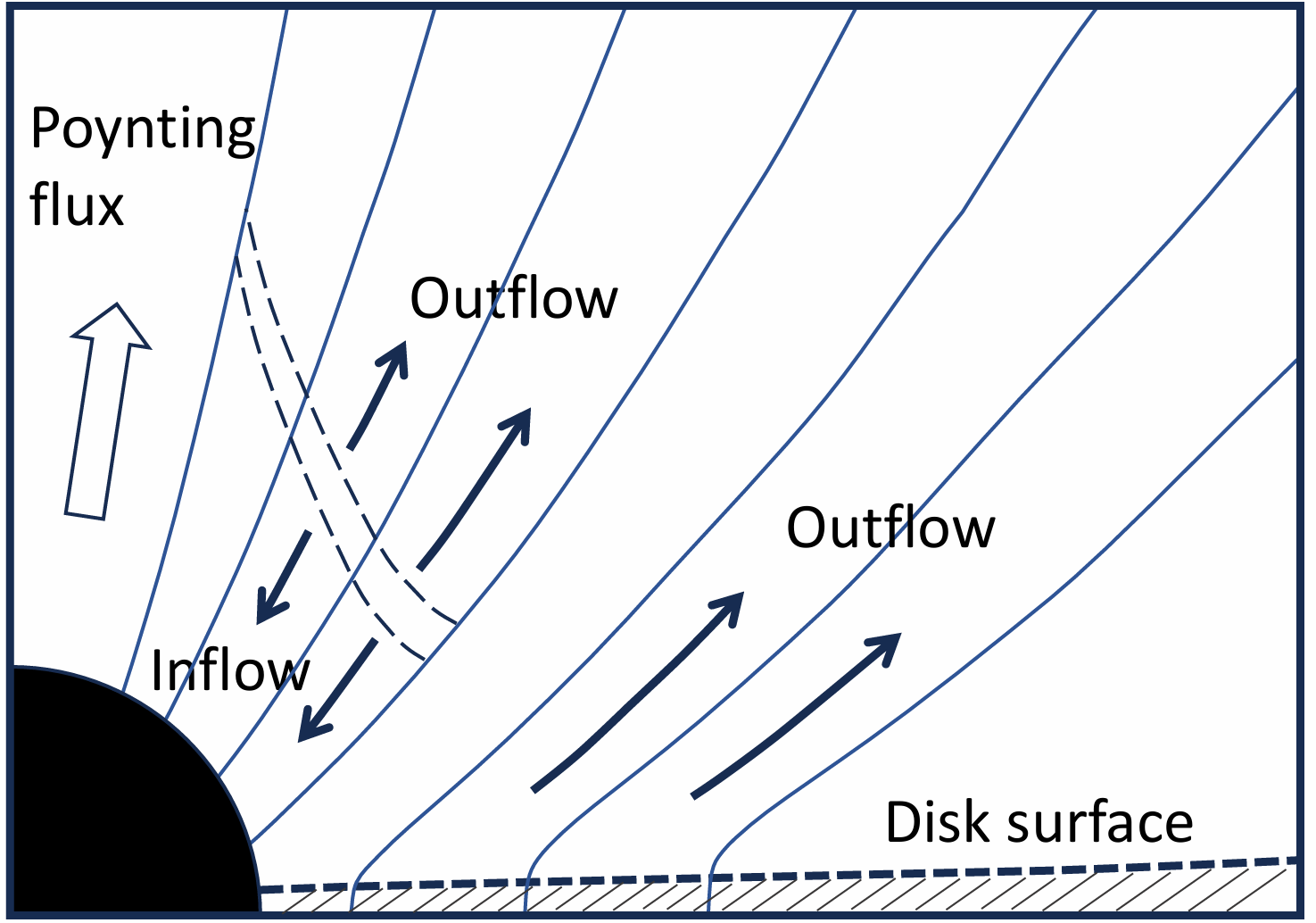}
	\centering
	\caption{Schematic illustration of the jet structure. The BZ process operates along magnetic field lines that thread the black hole horizon, which carry a continuous outward Poynting flux. Along these field lines, plasma inflows occur below the loading zone, while outflows are launched above it. Concurrently, the BP process is realized along magnetic field lines anchored in the accretion disk, which drive outflows via magnetocentrifugal acceleration and magnetic pressure gradients.}
	\label{fig:illu}
\end{figure}

Jet launching is generally attributed to the rotational dynamics of black hole accretion systems.
Horizon-scale magnetic fields required to sustain the jet are supplied by the disk: field lines anchored to its surface are advected inward and, once near the horizon, can be captured by the black hole. In this region, two established energy-extraction mechanisms operate: the Blandford-Znajek (BZ) and Blandford-Payne (BP) processes \cite{1977MNRAS.179..433B, 1982MNRAS.199..883B}. 
The BZ process extracts black hole rotational energy via the differential rotation between the horizon and the field lines threading it, producing outward Poynting flux. The BP mechanism accelerates plasma off the disk surface along field lines anchored in the disk when centrifugal forces overcome gravity \cite{pudritz1983centrifugally, ouyed1997numerical}.
Although the BZ process is often considered the dominant source of AGN jet power, both mechanisms likely operate simultaneously because magnetic field lines naturally couple to both the black hole and the disk \cite{Hawley:2005xs, McKinney:2006tf, komissarov2007magnetic, Tchekhovskoy:2011zx, McKinney:2012vh, Davelaar:2023dhl}, as is shown in Fig.~\ref{fig:illu}.

Outflowing matter exhibits distinct transverse structure. Centrifugal forces and low mass loading evacuate the polar region, forming a low-density, force-free spine in which electromagnetic energy is carried outward as Poynting flux and can accelerate particles to produce high-energy emission \cite{Blandford:2018iot,  Hirotani:2015fxp, Hirotani:2017ihv, Fendt:2003ke}.
Meanwhile, external pressure from the corona or disk winds confines outflowing plasma into a denser, matter-loaded sheath \cite{Hawley:2005xs,Moscibrodzka:2015pda,Globus:2016jtt, Nakamura:2018htq, Chatterjee:2019poz}.
The sheath retains substantial magnetic energy that efficiently accelerates the entrained plasma bulk flow, driving relativistic outflows and contributing predominantly to the millimeter-band emission \cite{EventHorizonTelescope:2019pgp,EventHorizonTelescope:2019pcy}; its non-negligible mass loading requires a full magnetohydrodynamics (MHD) treatment rather than a force-free approximation \cite{Huang:2019wqv, Huang:2020lvl}.
Typically, the BZ region encompasses both the spine and the sheath, whereas BP outflows launched from larger disk radii could only contribute to the sheath component \cite{Casse:2003mn,Hawley:2005xs}.

Near black holes, strong gravitational fields require a general relativistic magnetohydrodynamics (GRMHD) framework to model outflow dynamics \cite{Hawley:2005xs, McKinney:2006tf, komissarov2007magnetic, Tchekhovskoy:2011zx, McKinney:2012vh, Mizuno:2022vqa, Nakamura:2018htq, Narayan:2021qfw}. 
In this context, semi-analytical ideal MHD models offer essential physical insights and quantitative benchmarks \cite{1986A&A...162...32C, phinney1983theory, 1990ApJ...363..206T, 1991PASJ...43..569T, 1998PASJ...50..271T, Takahashi_2008, Pu:2017akw,Huang:2019wqv,Huang:2020lvl, Takahashi:2021xqb, Hou:2023bep, Azreg-Ainou:2024qqm, Hou:2024qqo, Tomimatsu:2003uz, Mitra:2022iiv, Kino:2022xme, Mitra:2024sqb}, where the plasma flow is steady and axisymmetric, following a frozen-in pattern in the poloidal plane. Within this framework, jet dynamics are primarily governed by two equations: the relativistic Bernoulli equation, describing energy conversion along field lines \cite{Pudritz:2006xf, Beskin:2010iba}, and the trans-field Grad–Shafranov (G-S) equation, determining the global magnetic field structure \cite{thorne1982electrodynamics, PhysRevD.44.2295, Gralla:2014yja,Chen:2020ipj, Pan:2017npg}.
Though the ideal MHD description fails to capture kinetic plasma processes in nearly collisionless regions or the regions with finite resistivity, it can still well describe the large-scale acceleration and collimation where dissipation processes are subdominant.

The semi-analytical modeling of ideal-MHD jet flows traces back to the solar-wind studies and the cold-outflow models from rotating neutron stars \cite{1967ApJ...148..217W, 1969ApJ...158..727M, 1970ApJ...160..971G}.  The ideal-MHD conservation laws were later generalized to a generic stationary, axisymmetric spacetime in \cite{PhysRevD.18.1809}.
Building on these foundations, \cite{1986A&A...156..137C, 1986A&A...162...32C, camenzind1987hydromagnetic} performed early GRMHD outflow analyses involving asymptotic critical points. 
\cite{1990ApJ...363..206T} first established the inflow–outflow structure and critical points of the poloidal flow equation in Kerr geometry, examined spin effects, and introduced the MHD Penrose mechanism. Follow-up studies refined the critical-point analysis \cite{1991PASJ...43..569T, 1998PASJ...50..271T}, explored the hot-flow behavior \cite{Takahashi_2002}, investigated the MHD Penrose process in different accretion states \cite{Pu:2012nt, Pu:2012va}, analyzed magnetically dominated inflow–outflow solutions \cite{Pu:2015rja}, and reformulated fast-magnetosonic-point regularity conditions \cite{Tomimatsu_2003, Takahashi:2008yg, Pu:2020pky}.
To probe the trans-field structure, the G-S equation in Kerr spacetime was formulated and employed to solve for density profiles near the stagnation region under simplified magnetic geometries
\cite{PhysRevD.44.2295}. 
In general, however, the complexity of the G-S equation precludes closed-form solutions, and practical applications therefore rely on numerical methods
\cite{1995A&A...300..791F, 1996A&A...313..591F, Pan:2017npg, Huang:2019wqv, Huang:2020lvl}.

Previous theoretical studies of GRMHD jet flows have primarily focused on black-hole-driven outflows associated with the BZ process, supported by magnetic fields threading the event horizon. However, the luminous jet sheath may instead be supported by field lines threading both the horizon and the adjacent disk surface; numerous works suggest that the bright outer edge of the jet is anchored at this horizon-disk interface (e.g., \cite{Nakamura:2018htq,Cattorini:2022tvx,Davelaar:2023dhl}).
In this context, investigating GRMHD jets powered by different launching mechanisms is essential for interpreting observational data, as model-data comparisons can constrain key physical parameters such as launching conditions, magnetic-field configurations, and black-hole spin.
Moreover, although the accretion disk produces most of the emission near the horizon, the bidirectional jet outflows exhibit strong Doppler boosting at larger radii and a well-ordered linear-polarization pattern, enabling their morphology to be clearly resolved by future high‑resolution observations  \cite{Ayzenberg:2023hfw,Johnson:2023ynn,Johnson:2024ttr}. 


In this work, we try to establish systematically a framework to investigate  steady, axisymmetric, ideal MHD systems relevant to energy conversion and flow acceleration in both black hole-driven and disk-driven jets. We derive a general launching condition that identifies the permissible launching positions for cold outflows, which in turn imposes a novel constraint on the structure of disk-threading magnetic fields near the equatorial plane.
Building on this foundation, we propose a simplified hybrid BZ and BP jet model under prescribed magnetic field configurations in Kerr spacetime. By solving the Bernoulli equation and performing a critical-point analysis, we obtain the flow structure along different types of field lines. Our results show that the interface between the BZ and BP components naturally gives rise to a velocity shear layer, accompanied by localized enhancements in both density and velocity—features that may offer a physical explanation for the observed limb-brightening in relativistic jets.

Our analysis is largely algebraic and relies solely on the conservation laws governing stationary, axisymmetric, ideal MHD flows. For simplicity, a geometrically thin plasma loading region is assumed to exist in the vicinity of the black hole. Additionally, the fast magnetosonic point is placed at infinity, such that the terminal Lorentz factor—or equivalently, the asymptotic magnetization—serves as the only free parameter along a given field line. All other physical quantities, such as the mass flux, density and magnetization, are uniquely determined by the Bernoulli equation in conjunction with regularity conditions at the MHD critical points.

The remaining parts of this paper are organized as follows. In Section \ref{Sec:GRMHDbasic} we review the basic framework, including the conserved quantities in ideal MHD, the wind equation, and the treatment of the asymptotic fast magnetosonic point.
In Section \ref{Sec:jetlaunching}, we focus on plasma loading and derive the launching conditions for both equatorial and off-equatorial configurations. In Section \ref{Sec:jets in Kerr}, we analyze jet solutions along prescribed magnetic field lines that thread either the black hole (Section \ref{Sec:horizonthreadingKerr}) or the accretion disk (Section \ref{Sec:diskthreadingKerr}) in the Kerr spacetime.
In Section \ref{Sec:Hybrid}, we propose the hybrid jet model and discuss its principal features in detail. We summarize our findings and provide further discussion in Section \ref{Sec:summary}.

\section{Jets in ideal MHD}\label{Sec:GRMHDbasic}

We begin with a non-self-gravitating magnetofluid in a general stationary and axisymmetric spacetime. 
The spacetime line element can be written in the following form:
\be\label{lineelement}
\mathrm{d}s^2 = g_{tt}\mathrm{d}t^2 + 2g_{t\phi}\mathrm{d}t\mathrm{d}\phi + g_{\phi \phi}\mathrm{d}\phi^2 +
g_{\mP\mP}\mathrm{d}\mP^2 \,.
\ee
Here and thereafter, we label the poloidal coordinates $r,\t$ by the index $\mP$. Note that for the Kerr spacetime, the above metric form refers to the Boyer-Lindquist (BL) spherical coordinates \cite{Boyer:1966qh}. In the Kerr–Schild (KS) spherical coordinates commonly employed in GRMHD simulations \cite{kerr1965new}, the metric contains a non-diagonal $t-r$, $\phi-r$ components and therefore does not conform to this form. 

\subsection{Stationary axisymmetric flow}\label{Sec:MHD}
The jet flow may consist of fully ionized plasma supported by baryons in the accretion flow, or of electron-positron pairs generated through high energy photon–photon collisions \cite{wardle1998electron}. In this work, we focus exclusively on the baryon-loaded case, in which the jet dynamics is governed by ions.
The energy-momentum tensor takes the form of $T^{\m\n} = T^{\m\n}_{\rm m} + T^{\m\n}_{\rm EM}$, where 
\bea
T^{\m\n}_{\rm m} = h u^{\m}u^{\n} + P \, g^{\m\n} \,, \quad
T^{\m\n}_{\rm EM} = F^{\m\rho}F^\n_\rho-\frac{1}{4}g^{\m\n}F^{\alpha \beta}F_{\alpha \beta} \, ,
\eea
represent the matter stress tensor and electromagnetic stress tensor, respectively; $h = e + P$ is the rest-frame enthalpy density; $e$, $P$ denote the rest-frame energy density and pressure, $u^{\m}$ is the flow four-velocity, $F^{\m\n}$ is the electromagnetic field tensor.
Throughout this work, we require infinite conductivity which is also known as ideal MHD condition, ensuring that the electric field measured by the plasma is zero, $u_\m F^{\m\n}=0$\footnote{Although finite resistivity effects—such as magnetic reconnection and current sheet formation—play a crucial role in understanding the details of magnetofluids, the ideal MHD approximation offers a well physical framework when the focus is on global fluid properties or time-averaged structures \cite{Mizuno:2022vqa}.}.

The evolution of the magnetized fluid is governed by the GRMHD equations, namely the conservation of energy-momentum: $\nabla_ \m T^{\m\n} = 0 $, and the conservation of particle number: $\nabla_\m(\rho u^{\m}) = 0 $, where $\rho$ denotes the rest-frame mass density. To close the system, an equation of state for the ion component is required. Under the adiabatic approximation, the total energy density is described by a polytropic relation $e = \rho + \left(\Gamma - 1\right)^{-1}P$, where $\Gamma$ is the adiabatic index, taking the value $\Gamma = 5/3$ for non-relativistic ions and $\Gamma = 4/3$ for relativistic ions.

To proceed, we define the ``pseudo-electromagnetic fields" as the contractions of $F^{\mu\nu}$ and its dual with the covector $dt$,
 $E^{\m} = -(dt)_{\n} F^{\m\n}$, $B^{\m} = (dt)_{\n} (\star F)^{\m\n}$. \footnote{The pseudo–magnetic field widely used in GRMHD studies differs from both the magnetic field measured by normal observers and that defined in the fluid rest frame. A detailed comparison, along with the explicit transformation relations, is presented in Appendix~\ref{FieldConventions}.}
Under stationary and axisymmetric ideal MHD, the poloidal magnetic field is aligned with the poloidal flow velocity, 
$B^{r} u^{\t} = B^{\t} u^{r}$, and the field components take
\bea
\quad B^{r} = \frac{\partial_\t\psi}{\sqrt{-g}}\,, \quad B^{\t} = -\frac{\partial_r\psi}{\sqrt{-g}}\,, \quad  B^{\p} =  (u^{\p} - \Omega_F u^t)\f{B^\mP}{u^\mP} \,, \label{psiOmega}
\eea
where $\sqrt{-g}$ is the metric determinant, $\psi \equiv A_{\p}(r,\t)$ is the stream function serving
as a label for magnetic field lines and plays a central role in describing the field geometry, and 
$\Omega_F  \equiv F_{tr}/F_{r\p} = F_{t\t}/F_{\t\p}$ is the field-line angular velocity, which is conserved along each magnetic field line. 
More details on $B^{\m}$ are included in Appendix~\ref{MagneticField}.

A nonzero field line rotation drives a generic evolution of ideal MHD flows, causing an initially arbitrary magnetic field near the black hole to gradually become toroidal at large distances. This behavior can be analyzed using a corotating observer with four-velocity $u_{\,\text{co}}^{\m} \propto \left(1,0,0,\Omega_F\right)$.
A characteristic surface, known as the light cylinder \cite{1969ApJ...157..869G}, is defined by the condition $u_{\,\text{co}}^2 \propto -k_0  \rightarrow 0$ where $k_0 = -(g_{\p\p}\Omega_F^2+2g_{t\p}\Omega_F+g_{tt})$, at which the corotating observer becomes null. 
Beyond the light cylinder, no timelike corotating observers exist, and the flow lags behind the rotating field lines, leading to a strong toroidal magnetic field in the fluid frame.

Moreover, it is well-established that the stationary and axisymmetric ideal MHD system further possesses three conserved quantities of the magnetofluid along each magnetic field line \cite{PhysRevD.18.1809,1986A&A...162...32C}. We denote them as the mass flux $\eta$, specific angular momentum $L$, and the specific energy $E$:
\be
\bag
\label{conservedquantities}
\eta  \equiv& \frac{\sqrt{-g} \rho u^{r}}{ F_{\theta\phi} } = \frac{\sqrt{-g} \rho u^{\theta}}{  F_{\phi r}} 
=  \frac{\rho u^{\mP} }{ B^{\mP}}\,,  \\
L \equiv& \, \mu u_{\phi} -  \frac{\kappa B^{\phi}}{\eta}\,,  \quad  E  \equiv -\mu u_{t} - \frac{\kappa B^{\phi} \Omega_F}{\eta} \,, \\
\eag
\ee
where $\m \equiv h/\rho$ denotes the specific enthalpy, and $\kappa \equiv g_{t\phi}^2 - g_{tt} g_{\phi\phi}$.
The existence of these conserved quantities $\{\Omega_F, \eta,E,L\}$ helps to simplify solving the flow structure, for a given stream function $\psi(r,\t)$. In Appendix~\ref{ConservationLaws}, we present a concise derivation of these MHD conserved quantities.

\subsection{The wind equation}\label{Sec:windeq}

The flow four-velocity can be solved by using the MHD conserved quantities. 
Firstly, with the definition of $E$ and $L$, the $t-$, $\p-$components of the four-velocity can be rewritten as
\be
\label{utupsi}
\bag
u_t = \f{E}{\m} \f{(g_{tt}+g_{t\p}\Omega_F)(1 - \Omega_F l) + M_A^2}{k_0 - M_A^2} \,,  \quad u_\p = \f{E}{\m} \f{(g_{t\p}+g_{\p\p}\Omega_F)(1 - \Omega_F l)-M_A^2l}{k_0 - M_A^2} \, ,
\eag
\ee
where $l \equiv L/E$ denotes the angular momentum density. Here, we have also introduced a dimensionless Alfvénic Mach number in the poloidal plane \cite{1986A&A...162...32C}:
\bea
\label{defMach}
M_A \equiv \sqrt{\f{\mu \rho u_p^2}{B_p^2}} = \sqrt{\f{\mu u_p|\eta|}{B_p}}\,,
\eea
where $u_{p} \equiv  \sqrt{u^\mP u_\mP} = \sqrt{u^r u_r + u^{\t}u_{\t}}$ and  $B_{p} \equiv  \sqrt{B^\mP B_\mP} = \sqrt{B^r B_r + B^{\t}B_{\t}}$ are the magnitudes of the  poloidal  velocity and poloidal  magnetic field, respectively. 
$M_A$ characterizes the ratio between the flow velocity and the propagation speed of magnetic disturbances in the poloidal plane 
\footnote{The second equality in Eq.~\eqref{defMach} follows from $u_p/B_p = |u_\mP/B_\mP|$, 
which holds only in a stationary, axisymmetric configuration where $B^{r}u^{\theta} = B^{\theta}u^{r}$ (see Appendix~\ref{MagneticField} for more details).
Since $\eta$ is signed to indicate the flow direction, we use $|\eta|$ in defining the Alfvénic Mach number to ensure it remains a positive-definite quantity.}.
Combining Eqs.~\eqref{utupsi} with Eqs.~\eqref{psiOmega}, the toroidal component of the magnetic field is given by
\bea\label{Bphi02}
B^{\phi}= \eta E \,\f{ (g_{tt} + g_{t\phi} \Omega_F) l + (g_{t\phi} + g_{\phi\phi} \Omega_F)  }{\kappa (k_0 - M_A^2) } \,.
\eea
By substituting  Eqs.~\eqref{utupsi} into the normalization condition $u_\m u^\m = -1$, 
one can derive the relativistic Bernoulli equation (also called the wind equation) that governs the evolution of the poloidal velocity along the field lines. 
Through ideal MHD condition, $u^{\t} = B^{\t}u^r/B^r$, $u^r = \pm u_p \left(g_{rr}+g_{\t\t}(B^\t)^2/(B^r)^2\right)^{-1/2}$, where the sign ``$\pm$'' corresponds to outflow/inflow. 
The wind equation can be written as 
\bea
u_{p}^2+1=\left(\frac{E}{\mu}\right)^2\frac{k_0k_2-2k_2M_A^2-k_4M_A^4}{(k_0-M_A^2)^2}\,,
\label{wind_eq}
\eea
where 
\bea
k_0 \equiv -\left(g_{\p\p}\Omega_F^2+2g_{t\p}\Omega_F+g_{tt}\right) \,, \quad 
k_2\equiv \left(1-\Omega_F l\right)^2\,, \quad
k_4\equiv -\f{g_{\p\p}+2g_{t\p}l+g_{tt}l^2}{\k}\,.
\eea
Once the magnetic field and conserved quantities $\{\Omega_F, \eta, E, L\}$ are known, $u_p$ (and $M_A$) is determined by solving Eq.~\eqref{wind_eq}, and $u^t$, $u^{\p}$ are determined by Eq.~\eqref{utupsi}. 
For plasma with finite temperature, the wind equation is complicated. 
For example, for adiabatic ions, the enthalpy density is given by 
$h = \rho + \Gamma (\Gamma - 1)^{-1}P$ with $P \propto \rho^\Gamma$. 
Since $\rho = |\eta| B_p/u_p$, the specific enthalpy $\mu = h/\rho$ is a function of $u_p$ along a given field line. 
Substituting the equation of state into Eq.~\eqref{wind_eq} leads to a non-integer power polynomial equation in $u_p$,  which can only be solved numerically \cite{camenzind1987hydromagnetic}.

\begin{figure}[ht!]
\hspace*{3.2cm}
	\includegraphics[width=3.7in]{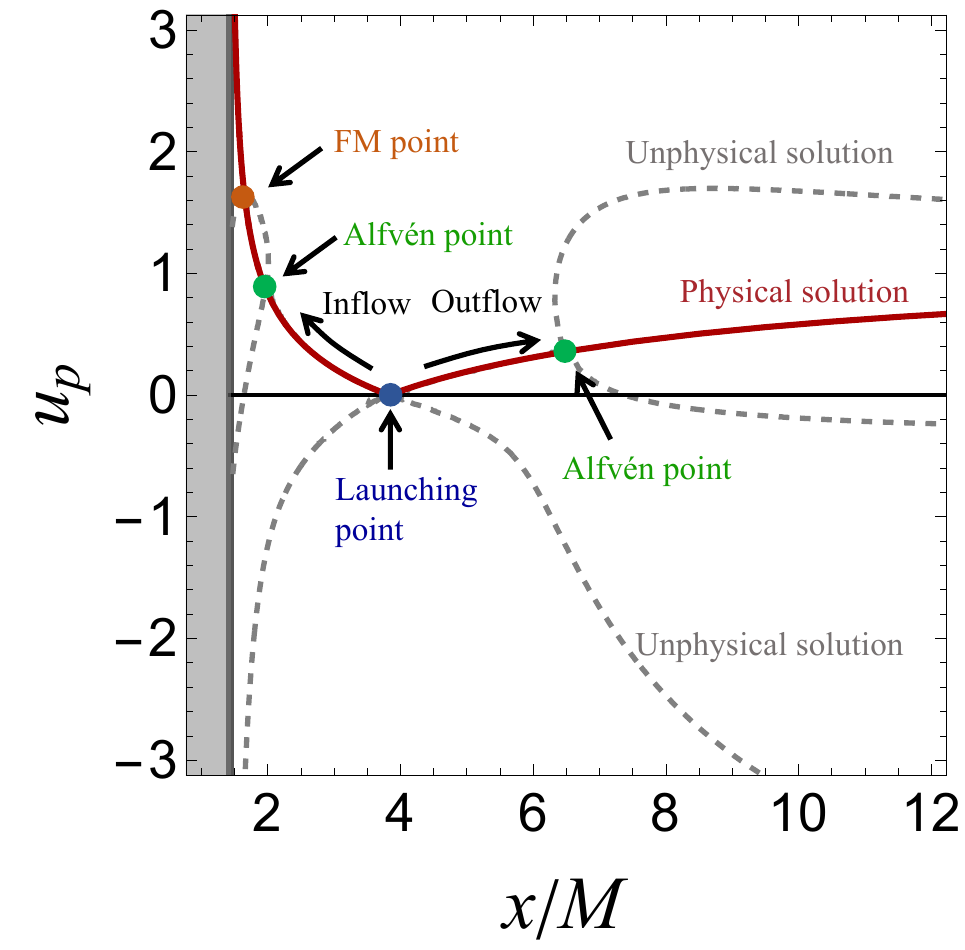}
	\caption{Real roots of the wind equation along a parabolic streamline in a cold MHD jet ($\m = 1$) in Kerr spacetime with 
	spin $a = 0.9$. The streamline is defined by $r-z = r_+$, where $r_+$ is the event horizon radius. Cylindrical coordinates are 
	given by $x = r \sin{\t}, z = r \cos{\t}$. 
	The conserved quantities are $\{\psi, \Omega_F, \eta, E, L\} = \{1.4359 , 0.1287, -0.2002, 0.6156, -0.4044 \}$ for the inflow and $\{ 1.4359, 0.1287, 0.03401, 3.375, 21.0432 \}$ for the outflow.  The red 
	solid curve shows the physical solution; gray dashed curves denote unphysical branches.}
	\label{Bz_up_a05}
\end{figure}

However, in the cold limit $\m \rightarrow 1$, substituting $M_A  = \sqrt{u_p|\eta|/B_p}$ into Eq.~\eqref{wind_eq} reduces it to a quartic equation in $u_p$. By analyzing the root behaviour of this equation, one can select the physically admissible solution of $u_p$. For a Kerr black hole, there are typically two real roots near the event horizon—one positive and one negative. Among them, only one root remains globally non-negative and smooth across the entire domain, as illustrated in Fig.~\ref{Bz_up_a05}.

To further characterize the solution structure, we need to examine the critical behavior of the flow at the points where the flow velocity reached characteristic MHD wave speeds \cite{1990ApJ...363..206T}.
Firstly, the denominator on the right-hand side of Eq.~\eqref{wind_eq} vanishes at the so-called Alfvén point, where $M_A^2 = k_0$. This point corresponds to the location where the poloidal velocity matches the Alfvén speed $u^2_A = B^2_p k_0 (\m \rho)^{-1}$, which characterizes the propagation speed of transverse magnetic disturbances along field lines.
Beyond this point, $u_p > u_A$, Alfvén waves are unable to propagate upstream. Mathematically, the Alfvén point constitutes a general critical point of the wind equation, where multiple roots of the equation coincide, as illustrated in Fig.~\ref{Bz_up_a05}. At this critical point, both the numerator and denominator of the wind equation simultaneously vanish~\cite{12467}. 

In addition, two other critical points commonly arise in the MHD flow: the slow magnetosonic (SM) point and the fast magnetosonic (FM) point. These points correspond to distinct dynamical transitions  and impose essential regularity conditions on the solution: the requirement that the flow passes smoothly through these critical points imposes constraints on the conserved quantities and plays a key role in selecting physically viable solutions. Further details regarding the properties of these MHD wave speeds and critical points are provided in Appendix~\ref{CriticalPoints}.

\subsection{Asymptotic magnetosonic point}\label{Sec:AMP}
In this work, we assume that the FM point of the jet outflow lies at spatial infinity, allowing FM waves to propagate freely throughout the jet interior. This treatment is equivalent to adopting the ``minimal-energy'' Michel-type solution \cite{1969ApJ...158..727M,1970ApJ...160..971G}. 
However, achieving more efficient magnetic-to-kinetic energy conversion and higher asymptotic outflow velocities requires placing the FM point at a finite location \cite{Pu:2015rja, Pu:2020pky}. 
The asymptotic-FM-point assumption used here thus represents a simplified limit and should be viewed as a conservative estimate of the achievable acceleration efficiency.
By imposing cold ions at the FM point at infinity, we can obtain the asymptotic expression for $E$ and $\eta$:
\bea
E = \gamma_{\infty}^3\,, \quad \eta = \f{\Omega_F^2  B_px^2 }{(\gamma_{\infty}^2 -1)^{3/2}}\Bigg|_{r \rightarrow \infty} \,.
\label{etainfity}
\eea
Here we adopt cylindrical coordinates defined by $x \equiv r \sin{\theta} $, $z \equiv r \cos{\theta} $. The quantity $\gamma_{\infty} = u^t\big|_{r \rightarrow \infty}$  denotes the Lorentz factor of the outflow as measured at infinity.
The first equation is the Michel's minimum energy \cite{1969ApJ...158..727M,1970ApJ...160..971G}, where the cubic dependence on $\gamma_{\infty}$ arises from the relativistic conversion between magnetic and kinetic energy, and the volume compression. 
Eqs.~\eqref{etainfity} are derived under the assumption of cold ions at infinity, without requiring the entire plasma flow to be cold. It provides an effective means to characterize plasma outflows along streamlines. Especially, for $u_p$ determined by the wind equation under a given magnetic field, the expression of $\eta$ in Eqs.~\eqref{etainfity} directly yields the ion number density $\rho = |\eta|B_p/u_p $.

We notice that the form of $\eta$ in Eqs.~\eqref{etainfity} actually implies opened magnetic field lines with the asymptotic behaviour of $B_p \sim 1/x^2$. 
The magnetic fields adopted in this work all satisfy such condition. Note that $d(B_p x^2)/dx < 0$ is required for efficient MHD acceleration and the existence of a finite FM point \cite{Tchekhovskoy:2009da}. As a result, for $B_p \sim 1/x^2$, the FM point is automatically pushed to infinity, and the system lacks a finite-radius transition where magnetic energy is fully converted into plasma kinetic energy.
This inefficiency can be reflected in the magnetization parameter, 
defined as the ratio of $E_{\rm EM} \equiv E - E_{\rm m}$ to $E_{\rm m} \equiv -\mu u_t$ \cite{Pu:2020pky},
\bea\label{SigmaM}
\sigma_M  \equiv  \f{E_{\rm EM}}{E_{\rm m}} = \f{\gamma_{\infty}^3 + \mu u_t}{ -\mu u_t} \,,
\eea
At the asymptotic FM point where the flow is cold, Eqs.~\eqref{utupsi} results in $u_t \rightarrow -\gamma_{\infty}$. Substituting this into Eq.~\eqref{SigmaM} yields $\sigma_M \big|_{r \rightarrow \infty}  \rightarrow  \gamma_{\infty}^2 - 1$, revealing that a substantial fraction of the energy remains stored in the electromagnetic field at infinity. This is a direct consequence of the field line geometry and the absence of a finite FM point. 
In this sense, $\sigma_M$ acts as a diagnostic of acceleration efficiency: high magnetization indicates a magnetically dominated jet at spatial infinity.

From the asymptotic analysis of the FM point, Eq.~\eqref{Bphi02} reduces to the following expression at large distances:
\be
\bag
B^{\p} \approx  -\,\f{\eta \left(\gamma_{\infty}^2-1\right) E }{\gamma_{\infty}^2 \Omega_F \k } \rightarrow
 -\f{\Omega_F \gamma_\infty B_p}{(\gamma_\infty^2 - 1)^{1/2}} \,.\label{Bphiinfty}
\eag
\ee
The asymptotic toroidal field magnitude is $\sqrt{B_{\p}B^{\p}} \sim \sqrt{g_{\p\p}} B_p \rightarrow x B_p $. Therefore, the magnetic field must take a toroidal shape at large distances. 
Under the limit $\gamma_{\infty} \gg 1$, we recover the result in the force-free electrodynamics (FFE), namely $\k \Omega_F B^{\p} \rightarrow - \eta E$ \cite{camenzind1987hydromagnetic}.

\section{Jet launching}\label{Sec:jetlaunching}

\subsection{The launching surface}\label{Sec:LS}

The MHD flow is loaded through a dense plasma layer near the black hole and driven outward by magnetic forces \cite{Huang:2019wqv,Chantry:2022ejm}. Here, we adopt a simplified model with a geometrically thin, axisymmetric launching surface (denoted by ``LS'') \cite{1990ApJ...363..206T}, where the poloidal outflow velocity is zero \cite{1986A&A...162...32C}.

Through Eqs.~\eqref{psiOmega}, the regularity of $B^{\phi}$ at the launching surface with $u_p|_{\rm LS} \rightarrow 0$ requires the plasma flow to be corotate with the magnetic field, $u^{\p}|_{\rm LS} = \Omega_F u^t|_{\rm LS}$.
From the definition Eqs.~\eqref{conservedquantities}, we obtain a constraint for $E,L$:
\bea
\label{ELstag}
\Omega_F L  = E - \m \sqrt{k_0}\Big|_{\rm LS}\,.
\eea 
Since the flow four-velocity is always timelike, the launching surface must lie within the light cylinder.

The constraint of the launching surface location can be found by expanding the flow-acceleration equation (or equivalently the MHD conservation law) near $u_p = 0$. 
The acceleration equation is obtained by differentiating the wind equation, Eq.~\eqref{wind_eq}, along the magnetic field line. It generally takes the form $u_p^{\prime} = N/D$,
where the prime ($^\prime$) denotes the gradient along the field line, $f^\prime = B^{\mP}\partial_\mP f$ for any scalar function $f$. The explicit expressions for $N,D$ are given in Appendix~\ref{Sec:FSpoint}. 
For a cold flow, 
expanding the acceleration equation in terms of $u_p$ yields
\bea\label{up_expend}
u_p^{\prime} \approx  k_0^{\prime}  \left( -\frac{B_p^{2} E^2 k_0 k_2}{2X}\f{1}{u_p} 
+ \frac{3 B_p E^4 k_2 (k_2 + k_0 k_4) \eta^3}{2 X^2}  + \mathcal{O}\left(u_p\right)  \right) \Bigg|_{\approx \rm LS} \,, 
\eea
where $X = B_p^{2} k_0^3 + E^2 (k_2 + k_0 k_4) \eta^2$.
Since Eq.~\eqref{ELstag}  implies $k_2 = k_0 E^{-2}$ and $k_0 > 0$ within the light cylinder, the launching surface must satisfy $k_0^{\prime} =0$. 
Moreover, if an outflow/inflow is to be launched from rest, $u_p^{\prime}$ must be positive/negative near the launching surface. Since $k_2 + k_0 k_4  = \kappa^{-1} \left[ g_{t\p} \left( 1 + l \Omega_F \right) + g_{tt}l + g_{\p\p} \Omega_F \right]^2 \geq 0$ always holds \cite{1990ApJ...363..206T}, we have the launching condition $ k_0^{\prime} \lessgtr 0$ for outflow/inflow, indicating that the launching surface should be the local maximum point of $k_0$. The constraint on the launching-surface location for thermal flows is presented in Appendix~\ref{Sec:appLC}.

The above discussion for the launching surface is universal. When we turn to different types of jets, the difference lies in the specific location of the launching surface and the boundary condition there. 
In our work, the BZ and BP jets are distinguished by the topology of their magnetic fields. Once the field geometry is chosen, the field angular velocity $\Omega_F$ is determined by the boundary conditions—either at the black hole horizon or at the disk. 

For simplicity, we adopt the cold limit in the following discussion. As demonstrated in previous studies, the ions within the accretion system in AGNs are largely non-relativistic, $k_B T_{\text{ion}} \ll m_{\text{ion}}$, so thermal effects do not significantly influence the jet dynamics \cite{Broderick:2010wi, Clausen-Brown:2011upb}.  
However, in the flows dominated by electron–positron pairs, the thermal energy can readily become comparable to the rest-mass energy, $k_B T_{\text{e}} \gtrsim m_{\text{e}}$, and the cold-fluid approximation is no longer valid \cite{beskin1992filling, wardle1998electron, Moscibrodzka:2011wj,  Broderick:2015swa, Hirotani:2015fxp, Hirotani:2016yyk, Hirotani:2017ihv, Chen:2018khs, Anantua:2019bna, Emami:2021ick, Yuan:2025war}.

\subsection{BZ process}\label{Sec:BZgeneral}
 
In the BZ jet, the magnetic field is captured by the black hole, and the plasma exhibit inward motion as it approaches the event horizon. Thus, there is a region that separates the inflow from the outflow, where the poloidal velocity goes to zero. 
In the ideal MHD framework, the separation is characterized by a two-dimensional surface and coincides with the launching surface \cite{Pu:2017akw, McKinney:2006tf}.
The differences in conserved quantities between the inflow and outflow satisfy \cite{Huang:2019wqv}
\bea
\begin{aligned}
(\eta_{\rm out} E_{\rm out} - \eta_{\rm in} E_{\rm in})  &= -\left(\eta_{\rm out} -  \eta_{\rm in} \right)  u_t \big |_{\rm LS}  \,,\\
(\eta_{\rm out} L_{\rm out} - \eta_{\rm in} L_{\rm in})  &=   \left(\eta_{\rm out} -  \eta_{\rm in} \right) u_{\p} \big|_{\rm LS} \,,\label{matchBZ}
\end{aligned}
\eea
where we use the subscript ``in'', ``out'' to label the quantities of the inflow and outflow, 
respectively\footnote{The equations in Eqs.~\eqref{matchBZ} are not independent; 
the second can be derived by combining the first with Eq.~\eqref{ELstag}.}.
The derivation of Eqs.~\eqref{matchBZ} is provided in Appendix~\ref{Sec:appLC}, where we assume purely particle-number loading for plasma injection within the loading zone. \cite{Huang:2019wqv}.
Therefore, $\eta$, $E$, and $L$ are conserved separately within the inflow and outflow, but are different across the launching surface. In contrast, the magnetic field and flow velocity remain continuous across the launching surface.

The acceleration of the outflow is easily determined by demanding an asymptotic FM point (see Sec.~\ref{Sec:AMP}). Specifically, $E_{\rm out}$, $\eta_{\rm out}$ are given by the asymptotic relation Eqs.~\eqref{etainfity}, and $L_{\rm out}$ is fixed by the relation Eq.~\eqref{ELstag} at the launching surface.
However, the inflow possesses finite FM point near the black hole, whose acceleration is more complicated. 
The existence of the inflow FM point can be understood as follows: the FM speed is the fastest propagation speed of disturbances in a magnetized fluid. Beyond the FM point, no disturbances can propagate from the downstream to the upstream. When sufficiently close to the event horizon, it is inevitable that no signal can propagate outward. Therefore, the FM point must occur outside the horizon (a detailed analysis can be found in \cite{1990ApJ...363..206T}).

To determine the inflow dynamics, we identify the FM‑point location together with the corresponding conserved quantities. This is done by combining 
(i) the wind equation, Eq.~\eqref{wind_eq}; 
(ii) the constraint on $E$ and $L$ at the launching surface, Eq.~\eqref{ELstag};
(iii) the matching conditions at the launching surface, Eqs.~\eqref{matchBZ};
(iv–v) the regularity conditions $D=0$ and $N=0$ imposed at the FM point through the acceleration equation $(\ln u_p)^{\prime} = N/D$ (Appendix~\ref{Sec:FSpoint}).

\subsection{BP process}\label{Sec:BPgeneral}

The BP jet originates from outflows launched at the surface of the accretion disk, where plasma is supplied along open or collimated magnetic field lines. For an ideal MHD accretion flow that is reflection-symmetric about the equatorial plane and exhibits vanishing polar velocity (i.e., zero velocity in the $\t$-direction), the following boundary conditions hold:
$B^r \big|_{\t = \pi/2} = E^\t \big|_{\t = \pi/2} = 0$. If a nonzero poloidal magnetic field component is present at the equatorial plane, i.e., $B^{\t} \big|_{\t = \pi/2} \neq 0$, then the field lines are frozen into the disk \cite{Hou:2024qqo}, implying that
\bea
\label{diskOmega}
\Omega_F  = \Omega_{\rm disk} \,,
\eea
where $\Omega_{\rm disk}$ is the angular velocity of the disk. In this case, the field lines are anchored in the disk and corotate with it.
In the cold MHD limit, the condition for jet launching is determined by $k_0^{\prime} = B^r \partial_r k_0 + B^\theta \partial_\theta k_0 = 0$.
Using $\Omega_F^{\prime} = 0$, this condition becomes
\bea\label{LSBP00}
k_0^{\prime} = -B^r \left( \pl_r g_{tt} + 2 \Omega_F \pl_r g_{t\p} + \Omega_F^2\pl_r g_{\p\p}  \right) - B^\t \left( \pl_\t g_{tt} + 2 \Omega_F \pl_\t g_{t\p} + \Omega_F^2\pl_\t g_{\p\p}  \right)  \,.
\eea
Interestingly, since the launching surface coincides with the disk surface, Eq.~\eqref{LSBP00} imposes a constraint on the disk height for a given $\psi$ and $\Omega_F$. Conversely, for a specified $\psi$ and disk height, Eq.~\eqref{LSBP00} determines the corresponding value of $\Omega_F$.  
Note that due to the reflection symmetry of the spacetime about the equatorial plane,  $\pl_{\t} g_{\mu\nu}\big|_{\t = \pi/2} = 0$. Hence, $k_0^{\prime} = 0$ is automatically satisfied at the equatorial plane, although this does not necessarily correspond to a local maximum of $k_0$.

For a geometrically thin disk, the jet launching surface lies close to the equatorial plane, and $\pl_{\t} g_{\mu\nu}\big|_{LS} \approx \pl_{\t} g_{\mu\nu}\big|_{\t = \pi/2} = 0$. In this regime, the launching condition simplifies to $0 = k_0^{\prime}\big|_{\rm LS} \approx -B^r \left( \pl_r g_{tt} + 2 \Omega_F \pl_r g_{t\p} + \Omega_F^2\pl_r g_{\p\p}  \right)\big|_{\rm LS}$.
Hence, if $B^r \big|_{\rm LS} = 0$, this condition is trivially satisfied. Otherwise, for $B^r \big|_{\rm LS} \neq 0$, solving the equation yields the angular velocity of the field-lines required for jet launching:
\bea\label{KepAngVel}
\Omega_F = \frac{-\partial_r g_{t\phi} + \sqrt{(\partial_r g_{t\phi})^2 - \partial_r g_{tt} \, \partial_r g_{\phi\phi}}}{\partial_r g_{\phi\phi}} \Bigg|_{\rm LS} \,,
\eea
which corresponds precisely to the Keplerian angular velocity.
In summary, for magnetic field lines threading a thin accretion disk, jet launching is permitted under two distinct conditions: (1) the magnetic field is perpendicular to both the equatorial plane and the disk surface, or (2) the disk rotates at the Keplerian angular velocity. To our knowledge, the result related to Eq.~\eqref{KepAngVel} has not been reported previously in the literature, and it applies to outflows launched from ideal-MHD, geometrically thin disks in stationary, axisymmetric spacetimes.

We note that GRMHD simulations favor sub-Keplerian flows yet still produce disk winds, a behavior that our current modeling fails to reproduce. This discrepancy likely stems from the simplified assumptions adopted here, i.e., a thin disk with negligible vertical velocity and a well-defined, near-equatorial launching surface for cold flows. By contrast, the simulated outflows are hot and primarily driven by magnetic-pressure gradients, often originating away from the equatorial plane and even lacking a clearly defined launching surface \cite{Dihingia:2021ncv, Davelaar:2023dhl}. 
Numerical resistivity also affects the field geometry, leading to possible discrepancies with semi-analytical predictions \cite{Nathanail:2024efu}. 
Besides, like most semi-analytical studies, our analysis assumes a stationary GRMHD solution, whereas time-averaged simulation results generally do not follow the GRMHD equations exactly.

\section{Specific models in the Kerr spacetime}\label{Sec:jets in Kerr}

The ideal MHD framework applies to general stationary, axisymmetric spacetimes. In this section, we focus on Kerr spacetime and examine jet flows under typical magnetic field configurations. Although the G-S equation is complex in curved spacetime \cite{PhysRevD.44.2295}, gravitational effects are significant only near the black hole \cite{Huang:2020lvl}. 
We therefore construct the magnetic field by adopting flat spacetime stream functions, and embedding them into the Kerr background through direct superposition.

\subsection{Parabolic magnetic fields}\label{Sec:horizonthreadingKerr}

\subsubsection{The field structure}\label{Sec:BZmag}

For the Kerr spacetime, we employ the BL coordinates $(t,r,\t,\p)$ and the cylinder coordinates defined as $x = r\sin{\t}$, $z = r\cos{\t}$. \footnote{Note that GRMHD simulations commonly employ KS coordinates. This gauge choice is not compatible with our theoretical analysis, which relies on the standard form of the line element in Eq.~\eqref{lineelement}.} For simplicity, we set the black hole mass to be 1. 
The horizon-threading magnetic field relevant to the BZ process can be well described by the following stream function \cite{Broderick:2008qf}:
\bea
\label{BZstream}
\psi(r, \theta) = C r^q(1- \cos{\theta}) \,,
\eea
where $0 < q < 2$ referred to as the collimation index, controls the asymptotic collimation of field lines. Increasing $q$ results in tighter collimation. The case $q = 0$ corresponds to a split-monopole configuration, whereas $q = 1$ produces a parabolic field structure.
The overall factor $C$ does not affect the jet dynamics and hereafter it will be set to 1.
Eq.~\eqref{BZstream} is in agreement with black hole magnetospheres in numerical simulations, where the typical collimation index is $q \simeq 0.75$ \cite{Tchekhovskoy:2008gq, McKinney:2006dx}. 

\begin{figure}[ht!]
\hspace*{-1cm}
	\includegraphics[width=7.4in]{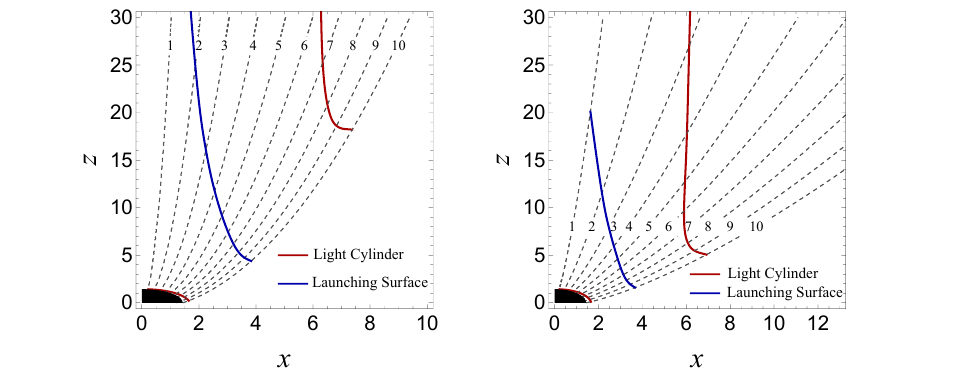}
	\caption{Poloidal magnetic field configurations corresponding to Eq.~\eqref{BZstream} are shown for $q = 1$ (left) and $q = 0.5$ (right). Gray dashed lines represent magnetic field lines anchored at various footpoint angles $\t_+ = n\pi/20$, $n = 1, 2, \dots, 10$; the associated values of $n$ are indicated in the figure. The blue curve denotes the jet launching surface, while red curves indicate the light cylinders.}
	\label{BZdiff_p}
\end{figure}

The stream function is defined in the northern hemisphere ($0\leq\t\leq\pi/2$); to ensure a vanishing net magnetic charge for the black hole, the form in the southern hemisphere is obtained by taking $\cos\theta \rightarrow -\cos\theta$ in Eq.~\eqref{BZstream}.
According to Eq.~\eqref{BZstream}, each field line is uniquely identified by its footpoint on the event horizon, 
labeled by the polar angle $\t_+(\psi) = \arccos{\left[1 - \left(r_+^{-q} \psi\right )\right]}$, so $\psi$ effectively encodes the angular position of the field-line at the event horizon.
In Fig.~\ref{BZdiff_p}, we present the poloidal magnetic field configurations with $q = 1$ (left panel) and $q = 0.5$ (right panel).
The gray dashed lines in each panel represent field lines with footpoints $\t_+ = n\pi/20$, $n = 1, 2, \dots, 10$.
 
The horizon-threading field lines must produce finite field components in any coordinate system that remains regular across the horizon. This requirement leads to the Znajek condition in the Kerr spacetime \cite{10.1093/mnras/179.3.457}:
\bea
\label{Znajek}
\Omega_F = \left[ \frac{a}{2r} + \frac{\k B^{\phi}}{2 r \sin^2{\theta}B^r} \right] \Bigg |_{r = r+}\,,
\eea
which serves as a constraint on the field-line angular velocity. As $B^{\phi}$ depends on $u^{\mu}$, Eq.~\eqref{Znajek} is highly coupled with the wind equation. 
Here, we rely on the FFE limit to obtain an explicit expression for $\Omega_F$. In the FFE regime, the energy is carried entirely by the electromagnetic field, giving $\eta_{\text{in}} E_{\text{in}} = - \Omega_F \kappa B^{\phi} \big|_{r = r+}$ by definitions (Eq.~\eqref{conservedquantities}). The inflow and outflow are related through $\eta_{\text{in}} E_{\text{in}} = \eta_{\text{out}} E_{\text{out}}$ \cite{Pu:2015rja}.
Under the asymptotic magnetosonic-point approximation, we have $E_{\text{out}} = \gamma_{\infty}^3$ and $\eta_{\text{out}} \approx \gamma_{\infty}^{-3}\Omega_F^2\left(B_px^2\right)\big|_{r \rightarrow \infty}$ as given in Eqs.~\eqref{etainfity}. This allows us to relate the toroidal field at the horizon to the poloidal field at infinity through
\bea\label{conn1}
\kappa B^{\p}\big|_{r = r+} \approx -\Omega_F\left(B_px^2\right)\big|_{r \rightarrow \infty}\,,
\eea
where the asymptotic field takes $B_p x^2\rightarrow  \sin^2\t_+$ for $q = 0$ and $B_px^2 \rightarrow  2\psi$ for $q > 0$.
Substituting Eq.~\eqref{conn1} into Eq.~\eqref{Znajek} yields the explicit expression of $\Omega_F$ for the stream function Eq.~\eqref{BZstream} :
\bea
\label{BZOmega}
\Omega_F =
\begin{cases}
\displaystyle  \frac{a}{4r_+ - a^2 \sin^2{\t_+}} \,, & \text{for } q = 0 \, ,  \\[1em]
\displaystyle  \frac{a \left(1+\cos{\t_+}\right)}{2(3r_+ + r_+\cos{\t_+}  - a^2 \sin^2{\t_+})} \,, & \text{for } 0 < q < 2 \,.
\end{cases}
\eea
It can be seen that for $q = 0$, $\Omega_F \geq \Omega_H/2$; whereas for $0 < q < 2$, $\Omega_F \leq \Omega_H/2$, where $\Omega_H = a(2 r_+)^{-1}$ is the black hole angular velocity. 
We emphasize that the FFE approximation is invoked only for obtaining the decoupled expression for $\Omega_F$  from Eq.~\eqref{Znajek}; the full MHD formulation is still used when computing the flow quantities along the field lines.

The black hole spacetime typically has an outer light cylinder far from the event horizon, and an inner light cylinder close to the event horizon \cite{1969ApJ...157..869G,okamoto1992evolution}, indicated by the red curves in Fig.~\ref{BZdiff_p} where $k_0 = 0$. In the region between the inner and outer light cylinders, $k_0 > 0$ always holds. It follows that the point where $k_0’ = 0$ must be a local maximum of $k_0$, ensuring a physical launching surface. 
The shapes of the launching surface and light cylinders reflect how the interplay between magnetic field rotation and spacetime rotation determine the field structures. 
In Appendix.~\ref{Sec:Kerrlightclinders}, we provide approximate expressions for light cylinders in the Kerr spacetime.

\subsubsection{The flow solution}\label{Sec:BZtypeflow}

\begin{figure}[ht!]
	\includegraphics[width=6.8in]{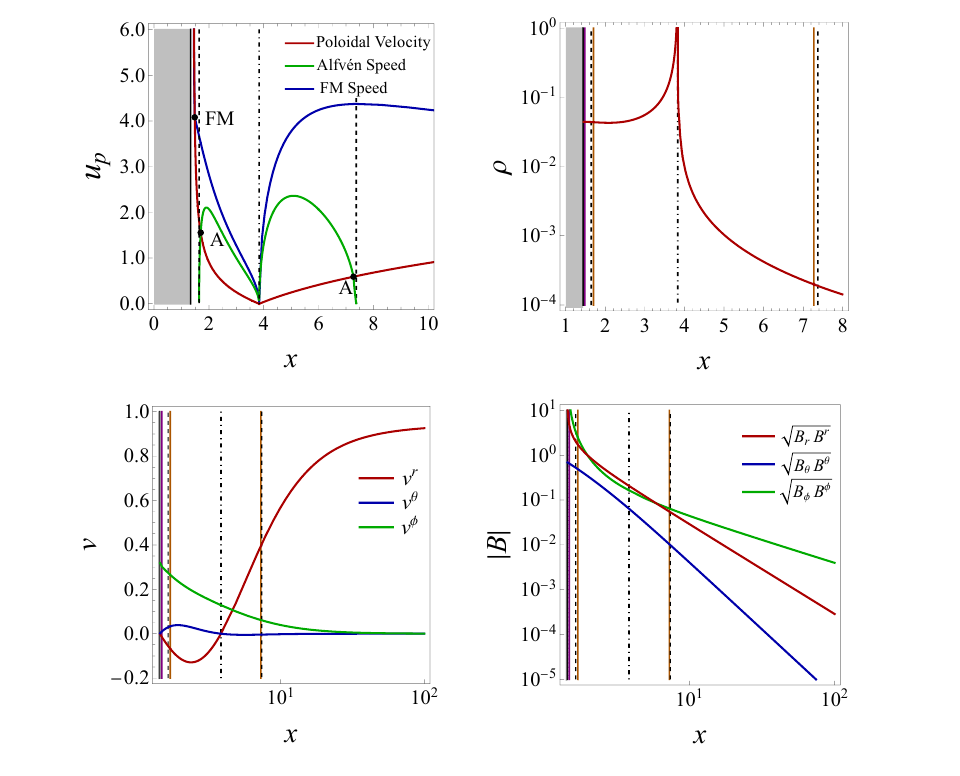}
	\caption{
	A representative BZ flow solution is shown, along a parabolic field line with $q = 1$, $\t_+ = \pi/2$. 
	In all panels, the gray-shaded region indicates the black hole interior; the black dot-dashed line marks the launching surface; black dashed lines show the light cylinders; the purple line marks the fast magnetosonic point; and orange lines mark the Alfvén points.
	\textbf{Top-left:} Poloidal velocity $u_p$ for both inflow and outflow (red), Alfvén speed (green), and fast magnetosonic speed (blue). The point labeled FM marks the fast magnetosonic point, while the points labeled A indicate the Alfvén points. 
	\textbf{Top-right:} Plasma number density. 
	\textbf{Bottom-left:} Components of the three-velocity: $v^r$, $v^\theta$, and $v^\phi$ where $v^{i} = u^{i}/u^t$.
	\textbf{Bottom-right:} Spatial components of the magnetic field.
	}
	\label{BZ_inout}
\end{figure}

Under the stream function Eq.~\eqref{BZstream}, the conserved mass flux $\eta$ for outflow along each field line is determined by the asymptotic expression Eqs.~\eqref{etainfity}, resulting in
\bea
\eta_{\rm out} = 
\begin{cases}
\displaystyle \frac{\psi \Omega_F^2 \left(1+\cos{\t_+}\right)}{(\gamma_\infty^2 - 1)^{3/2}}\,, & \text{for } q = 0 \,, \\[1em]
\displaystyle \frac{2\psi \Omega_F^2}{(\gamma_\infty^2 - 1)^{3/2}}\,, & \text{for } 0 < q < 2 \,.
\end{cases}
\label{eta_cases}
\eea
Combined with $E_{\rm out} = \gamma_\infty^3$ and $\Omega_F L_{\rm out}  = E_{\rm out} - \m \sqrt{k_0}\big|_{\rm LS}$, the wind equation Eq.~\eqref{wind_eq} can be solved to give $u_p = u_p(\eta_{\rm out}, E_{\rm out}, L_{\rm out}, r, \t)$. 
Then, $u_t$, $u_{\p}$ and $B^{\p}$ are determined through Eqs.~\eqref{utupsi} and Eq.~\eqref{Bphi02}, respectively. The outflow plasma density is given by $\rho = |\eta_{\rm out}|B_p/u_p $.
By using the matching condition Eqs.~\eqref{matchBZ} and the relation of $E$ and $\eta$ at the FM point, 
the inflow structure can be solved accordingly.
In Fig.~\ref{BZ_inout} presents a jet solution including both inflow and outflow, along a parabolic field line with $q = 1$, $\t_+ = \pi/2$.
We adopt a black hole spin of $a = 0.9$ and set the terminal Lorentz factor to $\gamma_\infty = 3$.
From these inputs, the conserved quantities of outflow are $\{E, L, \eta\}_{\rm out} = \{27,\ 204.6709,\ 0.0021\}$, and that of the inflow are $\{E, L, \eta\}_{\rm in} = \{-0.4134,\ -8.4023,\ -0.0382\}$.
The flow exhibits smooth transitions across the critical points.

As shown in the top-left panel of Fig.~\ref{BZ_inout}, the poloidal Alfvén speed, $u_A^2 = k_0B_p^2/\rho$, and poloidal FM speed, $u_F^2 = u_A^2 + \kappa \left(B^{\p}\right)^2/\rho$, increase from zero at the launching surface. 
Since Alfvén waves are transverse and propagate along magnetic field lines, their poloidal components diminish rapidly to zero in the regions near and far from the black hole, where the magnetic field becomes predominantly toroidal. 
In contrast, FM waves propagate in arbitrary directions rather than being constrained to the field lines. Consequently, the poloidal FM speed remains finite on both sides of the launching surface. 
 
The top-right panel of Fig.~\ref{BZ_inout} shows the plasma density. The outflow density drops rapidly with increasing $x$, while the inflow density exhibits modest variation near the black hole. The bottom-left panel presents the three-velocity components, $v^i = u^i/u^t$. In the outflow, $v^r$ dominates and approaches unity at large distances, while $v^\p$ decays to zero. In the inflow, frame dragging steadily enhances $v^\p$. The polar component $v^{\t}$ remains small throughout the flow.
The bottom-right panel displays the magnetic field components, which are continuous across the launching surface. At large distance, they asymptote to $\sqrt{B_r B^r} \rightarrow 2\psi / x^2$, $\sqrt{B_{\t} B^{\t}} \rightarrow 2\psi / x^3$ and $\sqrt{B_{\p} B^{\p}} \rightarrow 2\psi / x$, forming a highly toroidal structure with $|B^{\phi}| \gg |B^r| \gg |B^\theta|$.

Inflowing plasma with specific energy $E_{\rm in} < 0$ enables energy extraction from a rotating black hole via the MHD Penrose process \cite{Pu:2012nt,Toma:2024tan}, which fundamentally relies on the presence of the ergosphere. Specifically, when $|\Omega_H| > |\Omega_F| > 0$ and $\text{sign} \left(\Omega_H \right) = \text{sign}\left(\Omega_F \right)$, the existence of an Alfvén point within the ergosphere can lead to a negative inflow energy (see Appendix~\ref{App:Apoint}).

\subsection{Hyperbolic magnetic fields}\label{Sec:BPmag}\label{Sec:diskthreadingKerr}

For disk-anchored magnetic field lines, whether they collimate or open depends on the accretion system and the anchoring radius. Typically, inner-disk field lines tend to be more collimated, while outer-disk lines more readily open \cite{Fendt:2005fq, Gralla:2015vta, Dihingia:2021ncv}.
Because outflow launching and acceleration differ between these two geometries, they are best examined separately. In this section, we focus on outflows along open field lines, modeled with a simplified hyperbolic stream function. Collimated field lines are examined in Sec.~\ref{Sec:Hybridsetup}.
Up to an overall normalization, the hyperbolic stream function takes
\bea
\psi = \frac{x^2}{b^2+z^2}\,,\label{typeIs}
\eea
where $b$ is a constant. Constant $\psi,b$ parameterizes a hyperbolic field line within the poloidal plane, 
whose footpoint on the equatorial plane ($z = 0$) is given by $x_0(\psi) = b \sqrt{\psi}$. 
At large distances, the field line satisfies $x = \sqrt{\psi}z$, indicating that $\left(\psi\right)^{-1/2}$ serves as the asymptotic slope. In Appendix~\ref{TypeIIBP}, we examine an alternative hyperbolic field that arises as a solution to the FFE \cite{Gralla:2015vta, Chael:2023pwp}, and we observe similar outflow behavior to that discussed in this section.

\begin{figure}[ht!]
	\includegraphics[width=6.7in]{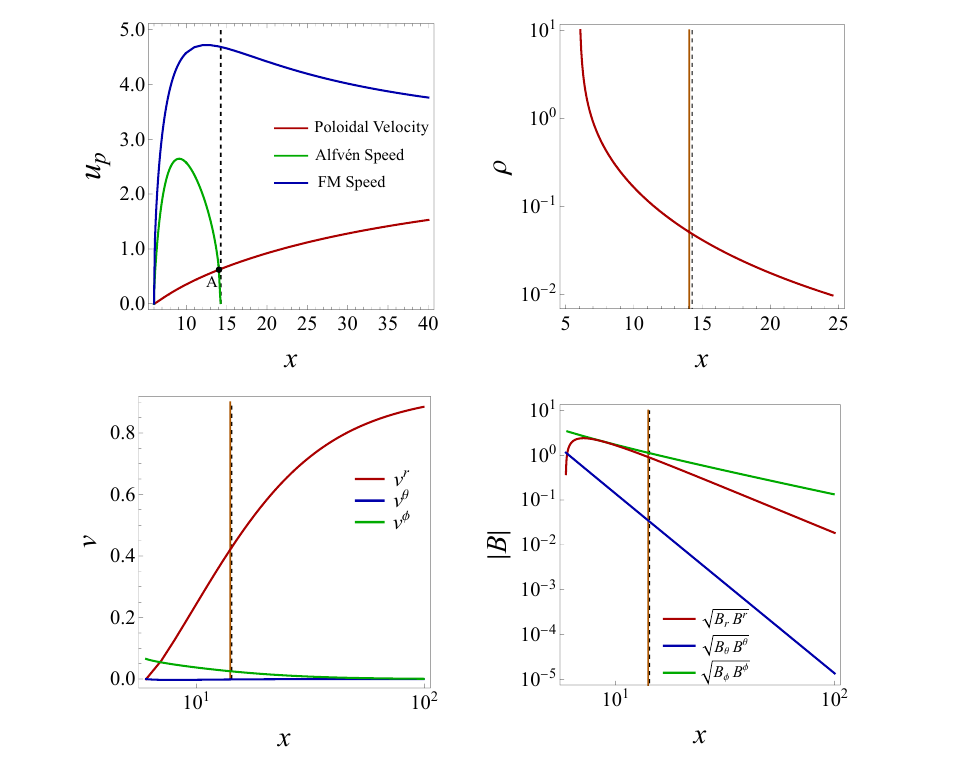}
	\caption{A representative BZ flow solution is shown, along a parabolic field line with $x_0 = (r_{\rm ISCO} - 1)\sqrt{20}$ and $b = r_{\rm ISCO} - 1$. We set the spin $a = 0.9$ and terminal Lorentz factor $\gamma_\infty = 3$. 
	The light cylinder is shown as black dashed lines, and the  Alfvén surface as an orange line.
	\textbf{Top-left:}  Poloidal velocity $u_p$ for outflow (red), Alfvén speed (green dashed), and FM speed (blue 
	dashed). The point labeled A indicates the outflow Alfvén points. 
	\textbf{Top-right}: Plasma density, which diverges near the launch point.
	\textbf{Bottom-left:}  Components of the three-velocity $v^r$, $v^\theta$, and $v^\phi$. 
	\textbf{Bottom-right:} Magnetic field components $\sqrt{B^r B_r}$, $\sqrt{B^\theta B_\theta}$, and $\sqrt{B^\phi B_\phi}$. }
	\label{BP1_out}
\end{figure}

As shown in Sec.~\ref{Sec:BPgeneral}, the angular velocity of the disk-threading field line is determined by the disk rotation itself. Here, we assume a thin Keplerian disk outside the innermost stable circular orbit (ISCO), with
$\Omega_{\rm disk} = \Omega_K = (x^{3/2} + a)^{-1}$, which determines the field-line angular velocity as
\bea
\label{BPOmega}
\Omega_F(\psi) = \frac{1}{b^{3/2}\psi^{3/4} + a} \,.
\eea
Furthermore, the asymptotic poloidal field is $B_p = 2\psi \sqrt{1+\psi}\,x^{-2} $. 
Applying the  asymptotic expression for $\eta$ (Eqs.~\eqref{etainfity}) to the hyperbolic stream function yields
\bea\label{etatype1}
\eta = \frac{ 2 \Omega_F^2 \psi\sqrt{\psi + 1}}{(\gamma_{\infty}^2 -1)^{3/2}} \,.
\eea
By combining Eqs.~\eqref{ELstag} and~\eqref{etatype1}, the wind equation \eqref{wind_eq} can be solved to obtain the flow velocity for given $\gamma_{\infty}$.
Fig.~\ref{BP1_out} illustrates the outflow structure, along a representative field line with $x_0 = \sqrt{20}(r_{\rm ISCO} - 1)$ and $b = r_{\rm ISCO} - 1$.
The top-left panel shows the poloidal flow velocity, Alfvén speed, and FM speed, all of which increase from zero at the disk surface. Beyond the Alfvén point, the poloidal Alfvén speed rapidly declines to zero, while the poloidal FM speed remains finite—analogous to the behavior observed in the BZ flow in Sec.~\ref{Sec:BZtypeflow}.

The radial component of the outflow clearly dominates and asymptotically approaches unity, while the azimuthal component gradually decays to zero, as shown in the bottom-left panel in Fig.~\ref{BP1_out}. The polar component remains negligible throughout the flow.
The bottom-right panel shows the spatial components of the magnetic field. At large radii, the field becomes strongly toroidal, characterized by: $\sqrt{B^r B_r}  \rightarrow 2\psi\sqrt{1+\psi}/x^2$,  $\sqrt{B^{\t} B_{\t}}  \rightarrow 2b^2\psi^{5/2}(1+\psi)^{-1}/x^4$ and $\sqrt{B^\phi B_\phi}  \rightarrow 2\psi \sqrt{1+\psi}/x$.

We briefly contrast our model with the classical BP scenario \cite{1982MNRAS.199..883B}, in which outflows are driven purely by the magneto‑centrifugal mechanism: rigid poloidal field lines transmit the centrifugal force, while the pressure and tension associated with the toroidal field component are neglected. In that case, outflows can be launched only if poloidal field lines emerge at angles less than $60^\circ$ to the disk surface \cite{1982MNRAS.199..883B, 1992ApJ...394..117P}. When general-relativistic effects are taken into account, such launching becomes easier in the vicinity of Kerr black holes \cite{Cao:1997tlg, Sadowski:2010aw}.
In contrast, Our ideal-MHD model includes both magnetic pressure and magnetic tension, predominantly supplied by the toroidal field component. The outflow is accelerated by the magnetic-pressure gradient, proportional to $ \sim \nabla (B^{\p})^2$, while the toroidal-field hoop stress governs its collimation. As shown in Sec.~\ref{Sec:Hybrid}, this enables launching from field lines with larger inclination angles.

For a disk-threading magnetic field line, solving $k_0’ = 0$ might yield two solutions: one exactly at the equatorial plane ($z = 0$), and the other slightly above it. The equatorial solution usually corresponds to a local minimum of 
$k_0$, where $B^r \big|_{\rm LS} = 0$ and both the magnetic-pressure gradient and magneto-centrifugal forces are negligible. In contrast, the off-equatorial solution corresponds to a local maximum of $k_0$ and is identified as the actual launching point, i.e., the disk surface. However, a subtle point arises: due to the geometrically thin disk assumption, only magnetic field lines with a local maximum of $k_0$ near the equatorial plane are admissible.

In the case of a hyperbolic field, the $k_0$ maximum always lies above the equatorial plane. For instance, under the parameters used in Fig.~\ref{BP1_out}, the solutions to $k_0’ = 0$ are $z =0$, $z = 0.2130$. The extent to which the launching surface deviates from the equatorial plane depends on the asymptotic slope of the hyperbola, as shown in the left panel of Fig.~\ref{slope}: increasing the slope shifts the $k_0$ maximum (marked by black dots) farther from the equatorial plane. A steeper slope yields more collimated field lines, requiring a greater vertical distance for $B^r$ to become sufficiently strong to drive an outflow. To remain consistent with the assumption of a geometrically thin disk, model parameters should be chosen such that the launching point remains close to the equatorial plane.

\section{Hybrid jet model}\label{Sec:Hybrid}

\subsection{Set up}\label{Sec:Hybridsetup}

In the preceding sections, we analyzed the flow structures along magnetic field lines that thread either the horizon or the disk independently. 
In realistic black hole magnetospheres, the global magnetic field is generated and sustained by the inflowing plasma. As the accretion flow advects inward toward the event horizon, it drags magnetic field lines along with it. These field lines are subsequently anchored to the spinning black hole on horizon scales. As a result, magnetic field lines can simultaneously thread both the event horizon and the accretion disk, enabling the coexistence of the BZ and BP processes \cite{Hawley:2005xs, McKinney:2006tf, komissarov2007magnetic, Tchekhovskoy:2011zx, McKinney:2012vh, Davelaar:2023dhl}.
It is evident that the magnetic field line structure must remain continuous across the boundary between the BZ and BP regions. More importantly, collimation is not limited to the BZ jet; it can also occur within the BP region near the horizon, driven by the rapid rotation of magnetic field lines \cite{McKinney:2012vh}.

In this section, we present a hybrid jet model that incorporates both horizon-threading and disk-threading magnetic field lines. To maintain consistency, we adopt the parabolic stream function introduced in Eq.~\eqref{BZstream}, which produces smoothly collimated field lines extending from the event horizon to larger distances: the field lines with $\psi \le r_+$ intersect the horizon and define the BZ region, while those with $\psi \ge r_+$ intersect the equatorial plane outside the horizon, corresponding to the BP region.
It is important to note that, as discussed in Sec.~\ref{Sec:BPgeneral}, magnetic field lines must be orthogonal to the equatorial plane. However, since our focus is on the flow just above the disk surface—slightly offset from the equatorial plane—the magnetic field can acquire a finite radial component, thereby enabling magneto-centrifugal acceleration (as illustrated in Fig.~\ref{fig:illu}). This justifies the use of the parabolic field configuration as an approximate representation of the BP region.

In the BZ region, we adopt the field-line angular velocity given in Eq.~\eqref{BZOmega}, which captures the interaction between the rotating black hole and the magnetic field. In the BP region, we assume $\Omega_F = \Omega_K$ outside the ISCO, allowing for jet launching near the equatorial plane, as discussed in Sec.~\ref{Sec:BPgeneral}.
We emphasize that this is a simplified prescription, based on the assumption of a geometrically thin, ideal MHD accretion disk with a correspondingly thin launching surface. In more realistic scenarios, sub-Keplerian flows are expected to develop near the equatorial plane. These flows are typically described in a phenomenological manner and will be investigated in future work.

\begin{figure}[ht!]
\hspace*{-7mm}
	\includegraphics[width=7.4in]{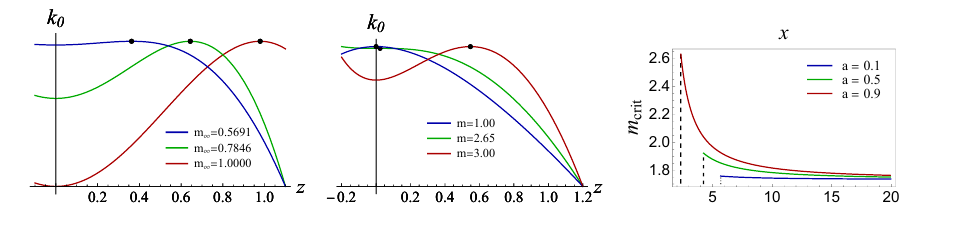}
	\caption{
		 \textbf{Left:} Variation of $k_0$ along disk-threading hyperbolic field lines [Eq.~\eqref{typeIs}], plotted for different asymptotic slopes $m_\infty \equiv \psi^{-1/2}$. 
\textbf{Middle:} Variation of $k_0$ along disk-threading parabolic field lines with various footpoint slopes $m$. In both panels, the footpoints of field lines are fixed at the ISCO, where $\Omega_F$ follows the Keplerian motion. For clarity, the values of $k_0$ have been rescaled in each plot to highlight the trends. 
\textbf{Right:} Critical slope for the parabolic field, above which the maximum of $k_0$ shifts away from the equatorial plane, shown as a function of the footpoint radius. The vertical dashed line marks the location of the ISCO.
	}
	\label{slope}
\end{figure}

For parabolic field configurations with $\Omega_F = \Omega_K$, introducing a nonzero slope at the equatorial plane allows for the existence of a near-equatorial local maximum of $k_0$—unlike the hyperbolic field case discussed 
in Sec.~\ref{Sec:diskthreadingKerr}. To illustrate this, we examine the variation of $k_0$ along field lines with different slopes. Specifically, we generalize the stream function in Eq.~\eqref{BZstream} to a parabolic form $z = \left(2\psi\right)^{-1}m(x^2 - \psi^2)$, where $m$ denotes the field line slope at the equator.\footnote{
The generalized form introduced here applies only to the parabolic case ($q = 1$) of Eq.~\eqref{BZstream}.
The parameter $m$ sets the equatorial slope and thus the opening of the parabolic field lines. Setting $m=1$ recovers Eq.~\eqref{BZstream}, while $m\neq1$ produces parabolic geometries with altered degrees of collimation.
}
The middle panel of Fig.~\ref{slope} shows how the slope $m$ affects the profile of $k_0$. 
Importantly, when the slope is less than a critical value, $m \leq m_{\rm crit}$, the maximum of $k_0$ remains anchored to the equatorial plane. The right panel of Fig.~\ref{slope} presents the dependence of $m_{\rm crit}$ on the footpoint radius of the field line. As the footpoint moves outward from the black hole, $m_{\rm crit}$ decreases; conversely, increasing the black hole spin leads to a higher value of $m_{\rm crit}$.
In this study, we focus on $m = 1$, which remains below the critical threshold for all spin values considered. As a result, the launching surface consistently resides near the equatorial plane across the entire parameter space.

In the region between the ISCO and the event horizon, plasma cannot maintain stable circular motion and instead plunges inward toward the black hole. In this region, the rotation of the magnetic field lines is expected to follow the angular velocity of the infalling plasma. Here, we assume that $\Omega_F$ equals the angular velocity of matter on a purely inspiraling trajectory originating from the ISCO, and set 
\bea
\Omega_F = \f{g^{\p\p}l_{\rm ISCO}-g^{t\p}}{g^{t\p}l_{\rm ISCO}-g^{tt}}\,,\quad  \text{for}\,\,\, r_+ \le \psi \le r_{\rm ISCO} \,,
\eea 
where $r_{\rm ISCO}$, $l_{\rm ISCO}$ denote the radius and angular momentum density at the ISCO, respectively.
Due to the absence of stable Keplerian orbits in this plunging region, the condition for a near-equatorial launching surface—required in Sec.~\ref{Sec:BPgeneral}—is no longer strictly satisfied. However, we find that for black hole spin  $a \gtrsim 0.75$, the deviation of the launching surface from the equatorial plane remains moderate\footnote{For example, at $a = 0.75$, the maximum vertical displacement of the launching surface occurs at the event horizon, where the local maximum of $k_0$ reaches $z = 1$. This deviation decreases rapidly with increasing black hole spin or radial distance from the horizon—at $a=0.9$, it reduces to $z=0.1$ at $r = r_+ + 0.2$.}. For spins $a \geq 0.9$, the deviation becomes negligible, allowing us to safely approximate the launching surface as near-equatorial and assign the associated conserved quantities accordingly.
Therefore, in the high-spin regime favored by jet observations, our model remains applicable and provides an effective description of the hybrid jet flow.

\subsection{Poloidal structures}\label{Sec:Hybridresult01}

\begin{figure}[ht!]
\hspace*{-6mm}
	\includegraphics[width=8in]{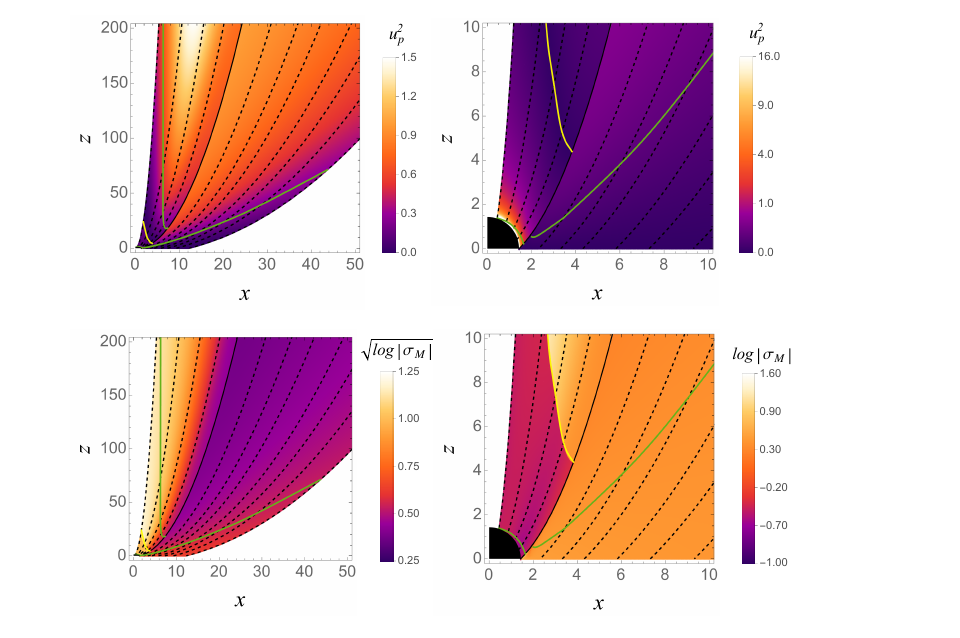}
	\caption{Distributions of the poloidal velocity (\textbf{Top}) and magnetization parameter (\textbf{Bottom}) for the hybrid model. The panels show the global structure (\textbf{Left}) and  zoom-in views (\textbf{Right}) toward the event horizon. In each panel, dashed lines trace magnetic field lines; the black solid line marks the BZ–BP interface; the yellow curve indicates the launching surface of BZ flows; and green curves denote the light cylinders.
	According to the definition $\sigma_M = \left(E + u_t\right)/(-u_t)$, the magnetization becomes negative in the inflow region; for clarity, we plot its absolute value. In the bottom-left panel, we omit the region with $|\sigma_M|<1$, as it appears only in the inflow and is negligibly small.
}
	\label{BZBP_up}
\end{figure}

\begin{figure}[ht!]
	\includegraphics[width=7in]{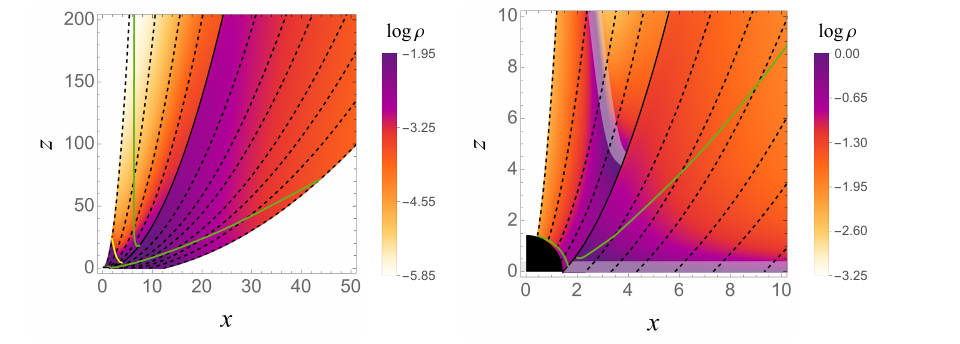}
	\caption{Plasma density distribution for the hybrid model. From left to right, the panels display the global structure and successive zoom-in views toward the event horizon. Gray-shaded bands are overlaid in the zoomed-in panel to indicate a manual cutoff near the launching surface, where the density formally diverges. All other plot elements are identical to those in Fig.~\ref{BZBP_up}.
	}
	\label{BZBP_rho}
\end{figure}

Employing the strategy for calculating MHD jet flows introduced in previous sections, we now turn to visual representations of the hybrid jet flow structure.
Fig.~\ref{BZBP_up} shows the distribution of the poloidal velocity $u_p$ and the magnetization parameter $\sigma_M = \left(E + u_t\right)/(-u_t)$ across magnetic field lines. Fig.~\ref{BZBP_rho} displays the corresponding density structure. In all figures, we set the black hole spin $a = 0.9$, the collimation index $q = 1$. 
Here, we employ a terminal Lorentz factor profile $\gamma_{\infty}(\psi) = 1.5 + 1.5 \exp\,\left[- (\frac{\psi}{0.3\sqrt{2}\, r_+} )^2 \right]$, representing efficient flow acceleration in the near-axis region. Other possible profiles of the terminal Lorentz factor are discussed in Sec.~\ref{Sec:Hybridresult02}.

We notice a clear difference in the shapes of the light cylinders between the BZ and BP flows, reflecting their distinct field-line rotations. Moreover, since the Alfvén point lies near the light cylinder, the transition from magnetically to kinetically dominated flow also occurs nearby, making the light cylinder a qualitative boundary between slow and fast flows \cite{Gelles_2025}. As shown in the left panels of Fig.~\ref{BZBP_up}, $u_p$ is lower and $|\sigma_M|$ higher below the light cylinder\footnote{By definition, $\sigma_M$ can become negative in the inflow region: for positive positive $(E_{\rm m})_{\text{in}} = -(u_t)_{\text{in}}$, one obtains $\sigma_{M}$ becomes negative if $E_{\text{in}} + (u_t)_{\text{in}} < 0$, corresponding to $\eta_{\text{in}}\kappa B^{\p}\Omega_F > 0$. This behavior simply arises from the inward matter flux ($\eta_{\text{in}} < 0$) and is unrelated to the conventional magnetization, $B^2/\rho$, which is always positive.}. 
This leads to a characteristic flow structure shaped by the light cylinder: slower speed and higher magnetization near the jet bases, and faster speed with lower magnetization near the BZ–BP interface. 

For the density profile, the hybrid jet possesses a higher density near the interface, which gradually decreases toward the inner/outer boundaries. Meanwhile, the density of the inflow is much higher than that of the outflow, which can be attributed to the strong gravitational capture of particles by the black hole within the loading region. However, we note that if a geometrically thick loading zone, rather than the thin launching surface assumed here, is taken into account, the density contrast between the inflow and outflow may be mitigated.

A distinct feature arises at the interface between the BZ and BP jets, where the poloidal velocity, magnetization, and density exhibit discontinuous behavior. This arises from a mismatch in the magnetic field rotations across the interface. Specifically, the rotation of horizon-threading field lines is governed primarily by the black-hole rotation itself, resulting in a characteristic relationship between the field-line angular velocity $\Omega_F$ and the black-hole angular velocity $\Omega_H$. For example, under the FFE approximation, efficient energy extraction requires $\Omega_F \approx \Omega_H/2$ in the BZ region \cite{1977MNRAS.179..433B}. 
Although the inertia of ions can retard the magnetic-field rotation, it is generally insufficient to overcome the black hole’s dominant influence—a conclusion supported by dynamical simulations \cite{Tchekhovskoy:2011zx,McKinney:2012vh}. In contrast, for field lines anchored in the accretion disk surface, they corotate with the accreting plasma, $\Omega_F = \Omega_{\text{disk}}$. As they approach the event horizon, the rotation of any equatorial disk configuration asymptotically matches that of the black hole, leading to $\Omega_F = \Omega_{\text{disk}} \rightarrow \Omega_H$ \cite{Hou:2024qqo}. Consequently, near the interface, the field-line rotation behaves as 
\bea
\Omega_F\big|_{\rm BP} \simeq  2\Omega_F\big|_{\rm BZ} \rightarrow \Omega_H \,, \quad \text{for} \,\,\, \psi \to r_+ \,.
\eea
In our hybrid model with spin parameter $a = 0.9$, the field line in the BZ region has a rotation rate of $\Omega_F \approx 0.1287$, while the adjacent BP field line exhibits a higher rotation rate of $\Omega_F \approx 0.3134$. It should be noted that at high spins, where the ISCO approaches the horizon, the magnetic field anchored in the plunging region appears as an outer sheath of the interface. The structures of the two are likely to be strongly correlated and left for future study.

The jump in the magnetic rotation arises largely from geometric effects and reflects the discontinuity between the BZ and BP regions. This feature likewise appears in dynamical accretion: the shear layer capable of energizing electrons forms along magnetic field lines anchored in the plunging region \cite{Davelaar:2023dhl}; thin-disk simulations even develop a third, turbulence-dominated outflow zone, distinct from the standard BZ and BP outflows \cite{Dihingia:2021ncv}, also located on field lines threading the plunging region. These suggest that our simplified model can already capture key characteristics of outflows in dynamical accretion systems.

\subsection{Asymptotic density profile}\label{Sec:Hybridresult02}

As a next step toward understanding the role of the BZ–BP hybrid, we turn to examine the asymptotic behavior of ion density.
Starting from the asymptotic expression for the mass flux Eq.~\eqref{etainfity}, the density far from the black hole can be approximated as
\bea
\rho\,\big|_{z \to \infty}  = \f{\eta B_p}{u_p}\, \approx \, \f{\Omega_F^2 B_p^2 x^2}{(\gamma_{\infty}^2-1)^{2}} \,.
\eea
Using the asymptotic scaling relations for a parabolic stream function, $B_p \approx 2\psi x^{-2}$, $z \approx (2\psi)^{-1} x^2$, we obtain the asymptotic expression for the density in terms of axial height as
$\rho \approx 2(\gamma_{\infty}^2 - 1)^{-2} \psi \, \Omega_F^2 z^{-1}$.
Accordingly, for a given height $z$, the field-line–dependent density distribution can be characterized by an effective density:
\bea\label{effrho}
\rho_{\rm eff} \equiv  z\rho\,\big|_{z \to \infty}  = \f{ 2 \psi\, \Omega_F^2}{(\gamma_{\infty}^2 - 1)^{2}} \,.
\eea
This relation indicates that field lines with higher angular velocities yield higher $\rho_{\rm eff}$, while those with larger terminal Lorentz factors correspond to lower $\rho_{\rm eff}$.
In realistic scenarios, $\gamma_{\infty}$ generally varies across field lines and is dynamically constrained by the trans-field G–S equation \cite{Huang:2019wqv,Huang:2020lvl}. The cross-field profile of $\gamma_{\infty}(\psi)$ thus plays a significant role in shaping both the velocity and density distributions in the poloidal plane. To qualitatively capture this effect, we adopt a synthetic profile modeled by a multipeak Gaussian function:
\bea\label{Gaussianprofile}
\gamma_{\infty}(\psi) =  \gamma_0 + \sum_{i = 1}^{\infty} \gamma_i \exp\left[ - \left( \f{\psi - \psi_i}{\d\psi_i} \right)^2 \right] \,,
\eea
where $\gamma_0$ sets the background Lorentz factor, and $\gamma_i$, $i \geq 1$ controls the amplitudes of the Gaussian peak centered at $\psi_i$ with characteristic width $\d\psi_i$.

\begin{figure}[ht!]
\hspace*{-8mm}
\includegraphics[width=7.3in]{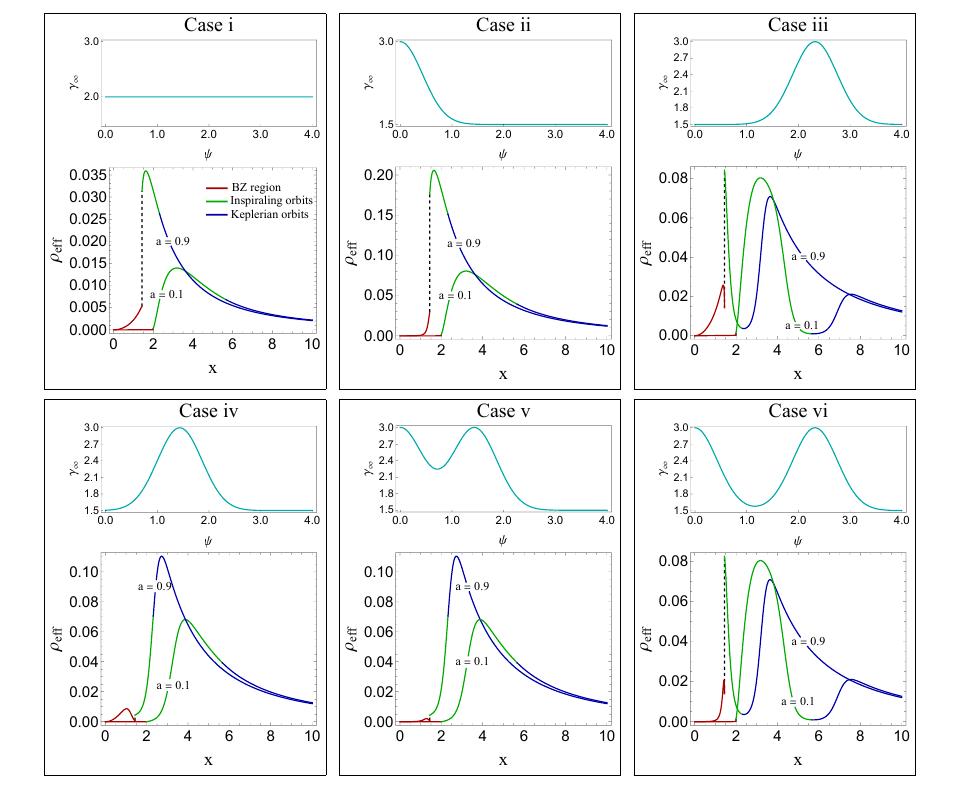}
	\caption{Effective density $\rho_{\rm eff}$ as a function of field-line footpoint coordinate $x$ is shown across the BZ region, the plunging region, and the Keplerian disk region. 
The top row and bottom-left panels display results for four representative single-peaked Lorentz factor profiles: 
	(i) uniform, (ii) polar-enhanced, (iii) ISCO-enhanced, and (iv) interface-enhanced. The bottom-middle and bottom-right panels present two double-peaked configurations: (v) polar plus interface-enhanced, and (vi) polar plus ISCO-enhanced. For each panel, the corresponding $\gamma_\infty(\psi)$ profile is displayed above it. }
	\label{rhoeff}
\end{figure}

We focus on four representative single-peaked cases with $\gamma_i = 0, i \geq 2$: 
(i) a baseline, uniform profile with $\gamma_0 = 2, \gamma_1 = 0$.
(ii) a polar-enhanced profile with $\gamma_0 =1.5, \gamma_1 =1.5$, $\psi_1 = 0, \d\psi_1 = 0.3\sqrt{2} r_+$, representing the typical situation in which the flow velocity is much higher in the BZ region.
(iii) an ISCO-enhanced profile with  $\gamma_0 =1.5,\gamma_1 =1.5$, $\psi_1 = r_{\rm ISCO}, \d\psi_1 = 0.3\sqrt{2}r_+$, concentrating the peak on field lines anchored at the ISCO.
(iv) an interface-enhanced profile with $\gamma_0 =1.5,\gamma_1 =1.5$, $\psi_1 = r_+, \d\psi_1 = 0.3\sqrt{2}r_+$, placing the peak at the BZ–BP interface.
The corresponding $\gamma_{\infty}(\psi)$ profiles for cases (i)–(iv) are presented in the upper subpanels of each panel in Fig.~\ref{rhoeff}. For the high-spin cases, cases (iii) and (iv) exhibit very similar behavior, as $r_{\rm ISCO} \to r_+$ in the limit $a \to 1$; these two cases likely capture a spine–sheath configuration with efficient acceleration occurring within the sheath \cite{Nakamura:2018htq}.

In the hybrid model considered here, a stream function used to label field lines is defined as follows: $\psi(\t_+) = r_+\left(1-\cos{\t_+}\right),\, 0 \leq \t_+ \leq \pi/2$ in the BZ region, and $\psi(x) = x,\, x \geq r_+$ in the BP region. To allow a unified horizontal axis across the interface, we extend $x$ into the BZ region through the substitution $x = r_+ \sin{\theta_+} \leq r_+$, thus covering horizon-threading field lines.
Fig.~\ref{rhoeff} shows the variation of the effective density $\rho_{\rm eff}$ across field lines, plotted against the footpoint coordinate $x$. 
For spin parameter $a = 0.9$, 
a clear discontinuity in $\rho_{\rm eff}$ is observed across the BZ–BP interface for all considered $\gamma_\infty$ profiles,
primarily due to the discontinuity in $\Omega_F$ here. 
Near the polar axis ($\psi \rightarrow 0$), $\rho_{\rm eff}$ approaches zero, indicating a quasi-vacuum jet spine induced by centrifugal forces. At larger $x$ (i.e., farther from the black hole), $\rho_{\rm eff}$ decreases monotonically, consistent with the density profiles shown in Fig.~\ref{BZBP_rho}. Regardless of the specific $\gamma_{\infty}$ profile, the density generally peaks in the inner disk region (i.e., $x \lesssim r_{\rm ISCO}$). However, the detailed structure of  $\rho_{\rm eff}$ remains sensitive to the particular shape of $\gamma_{\infty}(\psi)$.

Finally, we examine the influence of double-Gaussian-peak configurations by considering the following two cases: 
(v) a polar-enhanced plus interface-enhanced profile, characterized by 
$\gamma_0 = 1.5$, $\gamma_1 = 1.5$, $\gamma_2 = 1.5$, $\psi_1=0$, $\delta\psi_1 = 0.3\sqrt{2}r_+$, $\psi_2 = r_+$, 
and $\delta\psi_2 = 0.3\sqrt{2}r_+$;
(vi) a polar-enhanced plus ISCO-enhanced profile, with 
$\gamma_0 = 1.5$, $\gamma_1 = 1.5$, $\gamma_2 = 1.5$, $\psi_1=0$, $\delta\psi_1 = 0.3\sqrt{2}r_+$, $\psi_2 = r_{\rm ISCO}$, 
and $\delta\psi_2 = 0.3\sqrt{2}r_+$. 
The $\gamma_\infty(\psi)$ profiles for cases (v) and (vi) 
are shown in the upper subpanels of the bottom-middle and bottom-right panels in Fig.~\ref{rhoeff}.
The resulting effective density distributions are shown in the bottom-middle and bottom-right panels of Fig.~\ref{rhoeff}. 
In both cases, the polar-axis peak of $\gamma_{\infty}$ contributes little to the overall density profile, as it corresponds to a highly magnetized yet quasi-vacuum jet spine.

\section{Summary and discussion}\label{Sec:summary}

In this work, we developed a framework for steady, axisymmetric, ideal-MHD jets that encompasses both black hole–driven and disk-driven outflows, enabling semi-analytic exploration of jet launching, acceleration, and energy conversion. Starting from prescribed magnetic geometries, we solved the relativistic Bernoulli equation along each field line and performed a critical-point analysis. We derived a general criterion for launching cold outflows, applicable to a broad class of jets, and obtained a constraint on the rotation of disk-threading field lines required for near-equatorial outflow launching. 

Building on these results, we introduced a hybrid jet model based on a global parabolic magnetic configuration that links the BZ and BP regions. The model naturally produces a velocity shear and density discontinuity at the BZ–BP interface due to the jump in field‑line angular velocity, as illustrated through poloidal velocity and density profiles. We also showed that the light cylinder plays a central role in shaping the poloidal flow structure.

We analyzed how field-line angular velocity and terminal Lorentz factor shape jet structure through the effective density $\rho_{\rm eff} = 2\psi \Omega_F^2 / (\gamma_\infty^2 - 1)^2$.
Using synthetic multi-peaked Gaussian profiles for $\gamma_\infty(\psi)$, we found that $\rho_{\rm eff}$ consistently shows a strong discontinuity across the BZ–BP interface due to the jump in $\Omega_F$, along with a vacuum-like polar spine and a characteristic inner-disk peak. These features may help explain edge-brightened jets and transverse stratification. The sharp BZ–BP transition may also provide favorable conditions for MHD instabilities, including reconnection and associated non-thermal particle acceleration.

Despite its simplifying assumptions, our model reproduces key features of the black hole–disk–outflow system, including the inflow–outflow structure, the transition from magnetically to kinetically dominated regimes, and the sharp BZ–BP discontinuity. GRMHD simulations likewise show a jet–wind shear layer separating a highly relativistic jet from a mildly relativistic wind \cite{Davelaar:2023dhl}, and thin-disk simulations reveal a distinct BZ–BP transition marked by strong magnetic-pressure gradients and reconnection-driven plasmoids \cite{Dihingia:2021ncv}. Connecting this simulated transition region to the discontinuity in our model, and extending the framework to three-component outflows, remain promising avenues for future work.

Theoretically, the hybrid model can be extended by solving the G–S equation with appropriate boundary conditions to obtain a more realistic terminal Lorentz-factor profile $\gamma_\infty(\psi)$. 
Meanwhile, a more faithful treatment of jet launching further requires incorporating a finite-thickness mass-loading region. Moreover, because the light cylinder is a critical surface for jet dynamics, resolving the flow structure in its vicinity is also essential. Finally, relaxing the assumptions of a thin disk with zero vertical velocity and a near-equatorial launching surface would allow the model to be generalized to sub-Keplerian disks, enabling closer agreement with simulation results.

Recent advances in combining GRMHD with polarized radiative transfer have substantially deepened our understanding of jet imaging features \cite{Yao:2021phw,Cruz-Osorio:2021cob,Davelaar:2023dhl,Yang:2024kpz,Zhang:2024ddt}. The detection of synchrotron emission polarization and a ordered magnetic field in M87* \cite{EventHorizonTelescope:2021bee,EventHorizonTelescope:2021srq}, together with growing efforts in polarimetric jet modeling \cite{Chael:2023pwp,Gelles_2025,ALMA:2025wvr}, has further motivated the construction of jet models that link MHD theory to observables on scales of hundreds of gravitational radii \cite{Papoutsis:2022kzp,Zhang:2024lsf}.
The hybrid BZ–BP framework developed here captures the transition from the magnetically dominated jet base to the asymptotic outflow at large radii. The resulting jet morphology can be computed and used to produce synthetic images, enabling direct comparisons with GRMHD simulations and future high-resolution observations.

\section*{Acknowledgments}
We thank the anonymous referee for the valuable and expert feedback. We are grateful to Yosuke Mizuno, Indu K. Dihingia, Minyong Guo, and Zhenyu Zhang for insightful discussions.
The work is partly supported by NSFC Grant No. 12275004, 12588101. 
Y. H. acknowledges support from the NSFC Grant No. 12547123. 
L. H. acknowledges support from the NSFC Grant No. 12325302, 11933007 and the Shanghai Pilot Program for Basic Research, Chinese Academy of Sciences, Shanghai Branch (JCYJ-SHFY-2021-013).

\appendix

\section{MHD conservation laws}
In this section, we present a detailed analysis of the  stationary, axisymmetric magnetic field, and derive the explicit forms of the conserved quantities.

\subsection{Field transformation}\label{FieldConventions}

In the main text, we follow the convention commonly adopted in the literature and introduce the  ``pseudo-electromagnetic fields", which are defined as 
\bea\label{pseudoapp}
E^{\m} =  F^{t\m}  \,, \quad B^{\m} =   -(\star F)^{t\m} \,.
\eea
While these pseudo-fields do not directly represent the physical fields measured by any observer, they are widely used in GRMHD analyses because of their computational convenience.

A physical electromagnetic field is always defined with respect to a timelike observer. For an observer with four-velocity $U^{\mu}$, the electromagnetic fields measured in the comoving frame are
$\mathcal{E}^{\mu} = U_{\nu} F^{\mu\nu}$, 
$\mathcal{B}^{\mu} = -U_{\nu} (\star F)^{\mu\nu}$. 
The inverse relations then express $F_{\mu\nu}$ and its dual $\star F_{\mu\nu}$ in terms of the observer's four-velocity $U^{\mu}$ and the electric and magnetic fields
$\mathcal{E}^{\mu}$ and $\mathcal{B}^{\mu}$:
\be
\bag
&F_{\mu\nu}= U_{\m} \mathcal{E}_\n - U_\n \mathcal{E}_\m - \epsilon_{\m\n\a\b}\mathcal{B}^\a U^\b\,, \quad
(\star F)_{\mu\nu} = -U_{\m} \mathcal{B}_\n + U_\n \mathcal{B}_\m - \epsilon_{\m\n\a\b}\mathcal{E}^\a U^\b \,.
\eag
\ee
The transformation between the pseudo-fields and the physical fields 
$\mathcal{E}^{\mu}$, $\mathcal{B}^{\mu}$ is given by
\be
\label{trans}
\bag
 E^{\nu} = U^{t} \mathcal{E}^\n - U^\n \mathcal{E}^t - \epsilon^{t\n\a\b}\mathcal{B}_\a U_\b\,, \quad
B^{\nu}= U^{t} \mathcal{B}^\n - U^\n \mathcal{B}^t + \epsilon^{t\n\a\b}\mathcal{E}_\a U_\b \,,
\eag
\ee
As a special case, one may choose the normal observer with four-velocity
$n_{\mu} = (-\alpha,0,0,0)$, where $\alpha = (-g^{tt})^{-1/2}$ is the lapse function 
\cite{komissarov2004electrodynamics}. In this case, the relations reduce to $\mathcal{E}^{\mu} = \alpha E^{\mu}$, $\mathcal{B}^{\mu} = \alpha B^{\mu}$.

In MHD, the electromagnetic fields measured in the fluid frame are defined as $e^{\m} = u_{\n}F^{\m\n}$, $b^{\m} = -u_{\nu}(\star{F})^{\mu\nu}$, where $u^{\mu}$ denotes the fluid four-velocity.
For ideal MHD $e^{\m} = 0$,  the transformation between the pseudo-fields in Eq.~\eqref{pseudoapp}  and the comoving magnetic field takes the form
\bea\label{transBb}
&&E^{\m} = - \epsilon^{t\m\a\b}b_{\a}u_{\b} \,, \quad B^{\m} = b^{\m}u^{t} - b^{t}u^{\mu} \,, \\
&&b^{t} = B^{\m}u_{\m} \,, \quad  u^t b^i = B^i+ b^t u^i \,,
\eea
where $i$ represents the spatial indices. In addition, some studies (such as \cite{Takahashi_2008}) define the pseudo-electromagnetic fields as $E'_{\mu} = -F_{t\mu}$, $B'_{\mu} = (\star F)_{t\mu}$.
The relation between this convention and ours is given by
\bea\label{pseudorelation}
 E^{\nu} = g_{tt}^{-1}( -g^{\m\n} E'_\n  + g_{t\p} \epsilon^{t\n\a\p} B'_\a ) \,, 
 \quad
 B^{\nu} = g_{tt}^{-1}( - g^{\m\n} B'_\n  -  g_{t\p} \epsilon^{t\n\a\p} E'_\a ) \,.
\eea
This alternative definition is also adopted for computational convenience and does not correspond to physically measured fields. Note that neither $\{E^{\m},B^{\m}\}$ or $\{E'_{\m},B'_{\m}\}$ are genuine vectors, so the transformation in Eq.~\eqref{pseudorelation} cannot be obtained simply by raising or lowering indices.

\subsection{Stationary, axisymmetric magnetic field}
\label{MagneticField}
In a stationary and axisymmetric system, physical quantities vary only in the $(r,\t)$ plane---denoted as the poloidal plane. 
In this section, we would like to briefly review the derivation of the field structure by imposing symmetry requirements and the ideal MHD condition. 
Since the electromagnetic tensor only depends on $r,\t$, the azimuthal electric field vanishes, $E^{\p} = \k^{-1} \left( \pl_{\p} A_t - \pl_t A_{\p} \right) = 0$, where $\k \equiv g^2_{t\phi} - g_{tt}g_{\phi\phi}$. Employing the ideal MHD condition in the $t$ and $\phi$ directions, we obtain
\bea\label{poloidalEB}
E^{\mP}u_{\mP} = 0 \,, \quad B^{r} u^{\t} -  B^{\t} u^{r} = 0 \,. 
\eea
The first equation implies that the electric field is orthogonal to the plasma velocity in the poloidal plane. The second equation shows that the poloidal magnetic field is exactly aligned with the poloidal velocity, i.e., the magnetic field vector lies along the direction of fluid motion in the $(r, \theta)$ plane. This alignment is a direct consequence of the infinite conductivity assumption in ideal MHD.
The exceptional case is $u^r = u^{\t} = 0$, where the electromagnetic field cannot be determined by ideal MHD alone and additional conditions are required.   
Using the second equation in Eq.~\eqref{poloidalEB}, we have 
\bea\label{uA}
 B^{r} = \f{ \pl_{\t}A_{\p} }{\sqrt{-g}} \,, \quad B^{\t} = -\f{\pl_{r}A_{\p}}{\sqrt{-g}} \,, \quad u^\mP \pl_\mP A_\p =0 \,,
\eea
where $g$ is the determinant of the spacetime metric. We see that $A_\p$ is conserved along the fluid streamlines. In fact, $A_{\p}(r,\t)$ marks the magnetic flux enclosed by a circular loop of radius $r$ and polar angle $\t$, and commonly referred to as the stream function $\psi \equiv A_{\p}(r,\t)$.
The stream function serves as a label for magnetic field lines and plays a central role in describing the field geometry.

The ideal MHD condition in the poloidal directions leads to
\bea\label{Epoloidal}
&g_{rr} u^{t} E^r = g^{tt} u^{\t} F_{r\t} + \left( g^{tt}u^{\p} - g^{t\p}u^{t} \right) F_{r\p} \,, \nn \\
&g_{\t\t} u^{t} E^{\t} = g^{tt} u^{r} F_{\t r} + \left( g^{tt}u^{\p} - g^{t\p}u^{t} \right) F_{\t\p} \,.
\eea
Multiply the first equation of Eq.~\eqref{Epoloidal} by $B^r$, and the second equation by $B^{\t}$, then add the two resulting equations together to obtain
\bea\label{EBortho}
E^\mP B_\mP = 0 \,,
\eea
meaning that the pseudo-electric field and pseudo-magnetic field are orthogonal within the poloidal plane. Combining  Eq.~\eqref{EBortho} with the index lowering relation $g^{tt}F_{t\mP} = \left( g_{\mP\mP}E^\mP + g^{t\p}F_{\mP\p} \right)$, we can deduce  $F_{t\t}F_{r\p} = F_{tr}F_{\t\p}$. Hence, we can introduce the following field-line angular velocity:
\bea\label{defOmegaB}
\Omega_F \equiv \f{F_{t\mP}}{F_{\mP\p}} \,, \quad \mathcal{P} = r, \t \,,
\eea
representing the global rotation of the magnetic field line. 
By substituting Eq.~\eqref{defOmegaB} in the Bianchi identity in the $\p$ direction, we arrive at $F_{r\p} \pl_{\t}\Omega_F - F_{\t\p} \pl_{r}\Omega_F  + \left( \pl_{\t} F_{r\p} - \pl_r F_{\t\p}\right)\Omega_F  = 0$.
Since $F_{\mP\p} = \pl_{\mP}\psi$, the last term on the left-hand side of the equation vanishes automatically. Therefore, we have $B^\mP \pl_\mP \Omega_F = 0$, indicating that $\Omega_F$ is conserved along a field line within the poloidal plane. 

The field-line angular velocity is induced by the dragging from the rotating black hole and the accretion disk. For horizon-threading field lines, $\Omega_F$ is constrained by the regularity condition at the event horizon, known as the Znajek condition \cite{1977MNRAS.179..433B}. In contrast, for field lines anchored in the disk $\Omega_F$ is primarily determined by the local angular velocity of the accretion flow \cite{1982MNRAS.199..883B}.
Additionally, the pseudo-electric field can be rewritten as
\bea
E^{\mP} = - \left( g^{tt} g^{\mP\mP}\sqrt{-g}\right) \left(\Omega_F - \f{g^{t\p}}{g^{tt}} \right) \tilde{\epsilon}_{\mP\mP'} B^{\mP'} \,.
\eea
It becomes clear that the pseudo-electric field is induced by both the field line rotation and the spacetime frame dragging, represented by $g^{t\p}$.

\subsection{The MHD conserved quantities}
\label{ConservationLaws}

In stationary, axisymmetric ideal MHD, in addition to the stream function $\psi$ and field line angular velocity $\Omega_F$, the conservation of energy-momentum and particle number 
also yield three conserved quantities along field lines \cite{PhysRevD.18.1809}.  
Firstly, the ideal MHD condition in the $\phi$ direction yields:
$u^r F_{r\phi} + u^\theta F_{\theta\phi} = 0$. This allows us to define the mass flux $\eta$ as
\bea
\label{etadef}
\eta \equiv   \frac{\sqrt{-g} \rho u^{r}}{ F_{\theta\phi} } = \frac{\sqrt{-g} \rho u^{\theta}}{  F_{\phi r}} =  \frac{\rho u^{\mP} }{ B^{\mP}}  \,, 
\eea
which describes the particle number flux per unit magnetic flux.
Substituting Eq.\eqref{defOmegaB} into Eq.\eqref{etadef}, we obtain
$F_{tr} = -\sqrt{-g} \rho u^{\theta} \Omega_F \eta^{-1}$ and $F_{t\theta} = \sqrt{-g} \rho u^{r} \Omega_F\eta^{-1}$. Then, applying the Bianchi identity in the $\phi$ direction, $\pl_{r}F_{\t t} + \pl_{\t}F_{tr} = 0$, we have 
\bea
\partial_\mP \left(\sqrt{-g}\rho u^{\mP} \Omega_F \eta^{-1}\right)  = \sqrt{-g}\rho u^{\mP} \partial_\mP (\Omega_F \eta^{-1}) = 0 \,,
\eea
where we have used the particle number conservation, $\nabla_\mu (\rho u^\mu) = \partial_\mP \left( \sqrt{-g}\rho u^{\mP} \right)/\sqrt{-g}  = 0$. 
Since $\Omega_F$ is conserved along field lines, we have $B^\mP \partial_\mP \eta = u^\mP \partial_P \eta =  0$, which means that $\eta$ is also a conserved mass flux along field lines.
Furthermore, with the help of field-line angular velocity and mass flux, the energy conservation can be written as 
\bea\label{Econserv}
\bag
 \nabla_{\mu}T^{\mu}_{\,\,\ t} 
  = & \,\f{1}{\sqrt{-g}}\partial_\mP\left[\sqrt{-g}(h u^{\mP}u_{t} + F^{\mP\rho}F_{t\rho})\right]  \\
 = & \,\f{1} {\sqrt{-g}} \left [\pl_{\t}\psi \,\pl_r(\eta  \mu u_{t} +  \kappa B^{\phi}\Omega_F )   -  \pl_r\psi\,\partial_\theta(\eta  \mu u_{t} + \kappa B^{\phi}\Omega_F) \right ]  \\
 =  & \,B^\mP\partial_\mP \left  [\eta \left  (\mu u_{t} +  \frac{\kappa B^{\phi}\Omega_F}{\eta} \right )\right ] \,.
\eag
\eea
Through the same procedure, one can get a similar equation for the angular momentum conservation 
$\nabla_{\mu}T^{\mu}_{\,\,\ \p} = 0$. As a result, the specific energy and specific angular momentum
\bea
\label{defEL}
E \equiv -\mu u_{t} - \frac{\kappa B^{\phi} \Omega_F}{\eta} \,, \quad L  \equiv \mu u_{\phi} -  \frac{\kappa B^{\phi}}{\eta} 
\eea
are two conserved quantities along a field line, i.e., $B^\mP \partial_\mP E  = B^\mP \partial_\mP L = 0$.

Therefore, for a stationary and axisymmetric ideal MHD system, we have five conserved quantities:  $\{\psi, \Omega_F,\eta, E, L\}$, each of which is associated with a underlying symmetry of the MHD system. The stream function $\psi$ specifies the shape of magnetic field lines (as well as the streamlines) in the poloidal plane; $\Omega_F,\eta, E, L$ are invariant along each field line.

\section{Critical points}\label{CriticalPoints}

In Sec.~\ref{Sec:windeq}, we briefly outlined the existence of three critical points in the wind equation, and in Sec.~\ref{Sec:AMP}, we discussed how the asymptotic FM point constrains the conserved quantities. In this section, we provide a detailed analysis of the properties and roles of these critical points.

\subsection{The Alfvén point}\label{App:Apoint}

The Alfvén point is defined as the location where the denominator on the right-hand side of Eq.~\eqref{wind_eq} vanishes, marking the transition at which the poloidal plasma velocity equals the Alfvén speed:
\bea
u_p^2  =  u_{A}^2 \equiv \f{B^2_p k_0}{\m \rho}  \bigg|_{A} \,,
\eea
where the subscript ``A'' denotes the Alfvén point. Since the wind equation must remain regular at the Alfvén point, the numerator in Eq.~\eqref{wind_eq} must also vanish at this location:
\bea
\label{AlfvenCons}
(k_0k_2-2k_2M_A^2-k_4M_A^4)  \big|_{A} = -k_0(k_4k_0 + k_2) \big|_{A} = 0\,.
\eea
This condition determines the location of the Alfvén point along each streamline, forming a two-dimensional surface (a curve in the poloidal plane).
For a monopole-like magnetic field in flat spacetime, the Alfvén point is given by $x_A^2 = \Omega_F L/E$, which is often treated as a prescribed input parameter \cite{1986A&A...162...32C,1991PASJ...43..569T,1998PASJ...50..271T}.
In contrast, we do not fix the Alfvén point a priori. Instead, we specify the set of conserved quantities and require that the flow pass smoothly through all critical points.


The requirement that $u_t$ and $u_\phi$ (given by Eqs.~\eqref{utupsi}) remain regular at the Alfvén point, imposes the following constraint on the specific angular momentum:
\bea
l  = -\frac{g_{t\phi} + g_{\phi\phi} \Omega_F}{g_{tt} + g_{t\phi} \Omega_F} \bigg|_{A} \,,
\eea
indicating that the value of $l$ is fixed by the location of the Alfvén point.
Furthermore, the Alfvén point constrains the sign of the specific energy $E$ of the plasma. 
To clarify this, we express the specific energy in the form 
\bea
E = -k_0^{-1} (g_{tt} + \Omega_F g_{t\phi}) (E - \Omega_F L) \big|_{r_A} \,.
\eea
Given that $E - \Omega_F L = \sqrt{k_0}\big|_{\rm LS}$, the sign of $E$ determined by the sign of the metric combination $\left(g_{tt} + \Omega_F g_{t\phi}\right)\big|_{r_A}$.

The possibility of negative specific energy for inflowing plasma is intrinsically tied to the structure of the black hole ergosphere. In a rotating black hole spacetime, the ergosphere admits regions where $(g_{tt} + \Omega_F g_{t\phi})\big|_{r_A} > 0$, thereby allowing for $E < 0$. 
To make this explicit, we define a critical surface $r = r_{\rm C}(\t)$ at which $g_{tt} + \Omega_F g_{t\phi} = 0$. It can be shown that $r_{\rm C}(\t)$ always lies within the ergosphere (where $g_{tt} = 0$), and this critical surface exists only when $|\Omega_H| > |\Omega_F| > 0$ and $\text{sign} \left(\Omega_H \right) = \text{sign}\left(\Omega_F \right)$ \cite{1990ApJ...363..206T}. Therefore, if the Alfvén point lies inside the critical surface, then $\left(g_{tt} + \Omega_F g_{t\phi}\right)\big|_{r_A} > 0$, leading to a negative $E$ for the inflow. Such flows correspond to the extraction of energy from the black hole via the MHD Penrose process \cite{Pu:2012nt}.

\subsection{The acceleration equation}\label{Sec:FSpoint}
In addition to the Alfvén point, additional critical points can be identified by analyzing the flow acceleration equation.
Differentiating Eq.~\eqref{wind_eq} along the magnetic field lines yields 
\bea
(\ln{u_p})^{\prime} =  \frac{N}{D} = \frac{N}{D_{\rm A}D_{\rm SF}}\,,
\label{appacceleration}
\eea
where the prime ($^\prime$) denotes the gradiance along the field line, i.e., $f^\prime = B^{\mP}\partial_\mP f$ for a scalar function $f$. The functions appearing in the numerator and denominator are given by \cite{1990ApJ...363..206T}
\be
\label{N_D}
\bag
N =& \left[ -(1+u_p^2)(k_0 - M_A^2)^3 \frac{a_s^2}{1-a_s^2} + \left(\frac{E}{\mu} \right)^2 M_A^4(k_0k_4 + k_2)\right](\ln{B_p})^{\prime} \\
 & - \frac{1}{2(1-a_s^2)} \left(\frac{E}{\mu} \right)^2 [M_A^4(k_0 - M_A^2)k^{\prime}_4 + (k_0k_2-3k_2M_A^2 - 2k_4M_A^4)k_0^{\prime} ]\,,  \\
D_{\rm A} =& (k_0-M_A^2)^2 = (\frac{\m \rho}{B_p^2})^2(u_p^2-u_{A}^2)^2\,, \\
D_{\rm SF} =& \left[(u_p^2-c_s^2)(k_0-M_A^2)+\left(\frac{E}{\mu}\right)^2M_A^4\frac{(k_0k_4+k_2)}{(k_0-M_A^2)^2}\right] 
= -\frac{\m \rho}{B_p^2}(u_p^2-u_{FM}^2)(u_p^2-u_{SM}^2)\,.
\eag
\ee
Here, we have introduced the sound velocity $c_s$: 
\bea
c_s^2 = \frac{a_s^2}{1-a_s^2} \, , \quad a_s^2 = \frac{\partial \ln{h}}{\partial \ln{n}}\,,
\eea
where the partial derivatives are taken at fixed entropy, describing how the enthalpy responds to a logarithmic change in density under fixed-entropy conditions.
In Eq.~\eqref{appacceleration}, $D_{\rm A}D_{\rm SF} = 0$ determines three critical points, where $u_p$ takes one of the following characteristic speeds:
\bea
\bag
u_{A}^2 &\equiv \frac{B^2_p}{\m \rho}k_0\,, \\
u_{F}^2 &\equiv  \frac{1}{2}\left[\left(u_0^2+c_s^2\right)+\sqrt{\left(u_0^2+c_s^2\right)^2-4c_s^2u_{A}^2}\right]\,, \\
u_{S}^2 &\equiv  \frac{1}{2}\left[\left(u_0^2+c_s^2\right)-\sqrt{\left(u_0^2+c_s^2\right)^2-4c_s^2u_{A}^2}\right]\,,     
\label{Alf_FM_speed}           
\eag
\eea
representing the Alfvén speed, fast magnetosonic (FM) speed, and slow magnetosonic (SM) speed, respectively, where $u_0^2 = u_{A}^2+\f{\k (B^{\phi})^2 }{\mu \rho}$. The FM speed represents the maximum propagation speed of compressive MHD waves, incorporating both thermal and magnetic contributions. In contrast, the SM speed corresponds to the lower-speed branch of compressive modes.

Each critical point along the flow corresponds to a distinct dynamical transition. The SM point typically arises near the jet base (the loading zone) and marks the onset of acceleration, where the action of plasma pressure and magnetic forces overcomes gravitational confinement. The FM point, by contrast, defines the outermost critical surface; crossing this point signifies that the flow has become causally disconnected from the central engine and has reached its terminal acceleration regime.

The requirement of smooth transitions across critical points imposes regularity conditions, which translate into algebraic constraints on the conserved quantities of the flow. In the cold plasma limit ($c_s^2 = 0$), the FM speed reduces to $u_F = u_0$ and the SM speed becomes zero. 
Thus, the smoothness condition at the SM point is set by the boundary conditions at the launching surface. In contrast, at the FM point, smooth crossing requires both the numerator and denominator of Eq.~\eqref{appacceleration} to vanish simultaneously. This selects the physically admissible solutions.

\subsection{The asymptotic FM point}\label{appFMP}

At the FM point, 
by combining the wind equation Eq.~\eqref{wind_eq} with the condition $u_p = u_{F}$, we obtain:
\bea
\label{finE}
E^2 &=& \frac{\m^2(1+c_s^2)(k_0 - M_A^2)^3}{k_0^2 k_2 - 3 k_0 k_2 M_A^2 + 3 k_2 M_A^4 + k_4 M_A^6} \,\Bigg |_F\,, \\
\eta^2 &=& -\f{B_p^2}{\m^2}\left(-\frac{c_s^2}{M_A^4} +\frac{(1+c_s^2)(k_0 k_4 + k_2)}{k_0^2 k_2 - 3 k_0 k_2 M_A^2 + 3 k_2 M_A^4 + k_4 M_A^6}\right)^{-1} \Bigg |_F\,, \label{fineta}
\eea
where the Mach number reduces to $M_A^2 \rightarrow \mu |\eta| u_{F}B_p^{-1} \big |_F$.
Together with the regularity condition $N|_F = 0$, these equations constrain both the location of the FM point and the values of $E$ and $\eta$.

To characterize the outflow at large distances, we adopt cylindrical coordinates defined by $x \equiv r \sin\theta$, $z \equiv r \cos\theta$.
We assume a cold  plasma at infinity, such that $P = 0$, $\mu = 1$, and $c_s^2 = 0$.
Under this condition, the Mach number at infinity ($x \to \infty$) becomes
\bea
\label{inftyM}
M_A^2 \bigg |_{F}= k_0 + \k \f{(B^{\phi})^2}{B^2_p} \,\rightarrow\, - x^2 \Omega_F^2 + \f{x^2 \Omega_F^2}{v_{\infty}^2}= \f{x^2 \Omega_F^2 }{\gamma_{\infty}^2-1}\,,
\eea
where $v_{\infty}$ is the asymptotic, spatial poloidal velocity, and $\gamma_{\infty} = (1-v_{\infty}^2)^{-1/2}$ is the terminal Lorentz factor of the outflow.
By expanding Eq.~\eqref{finE} and Eq.~\eqref{fineta} in the limit of large $x$ and substituting Eq.\eqref{inftyM}, we derive the asymptotic forms of $E$ and $\eta$ as given in Eq.~\eqref{etainfity}.
The asymptotic expression for $B^{\p}$ can be derived by substituting Eqs.~\eqref{inftyM}~\eqref{etainfity} into the large distance limit of Eq.~\eqref{Bphi02}.

\section{Launching Condition}\label{Sec:appLC}

\subsection{The plasma loading}

The MHD flow is launched from a mass-loading zone \cite{Huang:2019wqv,Chantry:2022ejm}. 
Within the loading zone, there is a source term for plasma injection, $\nabla_{\mu}T^{\mu\nu} = S^{\nu}$. 
We consider that during the loading process, there is only particle number loading \cite{Huang:2019wqv}, resulting in
 $S^{\mu} = \eta^{\prime} \mu u^\mu$. 
Thus, according to Eq.~\eqref{Econserv}, the energy and angular momentum  transport equations are written as
\bea\label{loadingzoneTmn}
\nabla_{\mu}T^{\mu}_{\,\, \ t}  = -(\eta E)^{\prime}=  \eta^{\prime} \mu u_t \, , \quad \nabla_{\mu}T^{\mu}_{\,\, \ \phi}  = (\eta L)^{\prime} = \eta^{\prime} \mu u_{\p} \,. 
\eea

In the main text, we adopt a geometrically thin, axisymmetric launching surface with a zero poloidal outflow velocity \cite{1986A&A...162...32C}.
At the launching surface, the mass density $\rho = |\eta|B_p/u_p $ diverges due to vanishing $u_p$, while $\eta$, $B_p$ remain finite. This singularity can be regularized by introducing a finite-width loading zone, where plasma is supplied either from the external accretion flow or via pair production within the jet \cite{wardle1998electron, Globus:2014eaa}. However, modeling such processes lies beyond the scope of this study.

According to Eq.~\eqref{loadingzoneTmn}, there is a jump in the conserved quantities across the launching surface: 
\bea
\label{match}
\delta(\eta E) =\delta \eta (-\mu u_t) \,, \quad \delta(\eta L) =\delta\eta (\mu u_\phi) \, ,
\eea
where $\delta(X) = X_{\rm flow}-X_{\rm loading}\,$, $X \in \{ \eta E, \eta L, \eta\}$.  
Note that the limit $S_t \rightarrow 0$ gives a special case of an electromagnetically dominated jet with the matching condition $(\eta E)|_{\rm in} = (\eta E)|_{\rm out}$, as employed in \cite{Pu:2015rja}. 
In this work, we adopt the general formulation of Eqs.\eqref{match}, which incorporates plasma injection in the loading zone. The details within the loading zone are beyond the scope of this study. Instead, Eq.~\eqref{match} serves solely to provide a matching condition: for field lines threading the horizon, it leads to the relation between the conserved quantities of the inflow and outflow, as given in Eq.~\eqref{matchBZ}.

\subsection{The launching condition}

We derive the launching condition by analyzing the plasma acceleration near the launching surface. From energy-momentum conservation, we have $\nabla_{\mu}T_{\rm m}^{\mu\nu} = - \nabla_{\mu}T_{\rm EM}^{\mu\nu} = f^{\n}$, where $f^{\m}$ denotes the Lorentz force density. In ideal MHD, where the electric field vanishes in the comoving frame, $f^{\m}$ is orthogonal to the comoving magnetic field,  leading to
\bea\label{LorentzTmn}
b^\m f_{\m} = b_{\nu}\nabla_{\mu} ( h u^{\m}u^{\n} + P \, g^{\m\n})  =  0 \,.
\eea
Using $b^{\nu} u_{\nu} = 0$, this can be rewritten as $h\, b_{\nu} u^{\m} \nabla_{\m} u^{\n} + b_{\m} \partial^{\m} P = 0$.
Near the launching surface, the plasma motion is predominantly toroidal, with $u^{\p} \approx \Omega_F u^t$ and $u_p \approx 0$.
The derivative term $b_{\nu} u^{\m} \nabla_{\m} u^{\n}$ expanded in powers of $u_p$ as (also see Appendix A.4 in \cite{Gelles_2025})
\bea\label{derivativeterm}
b_{\nu} u^{\m} \nabla_{\m} u^{\n} \approx  B^{\mP} \partial_{\mP} \left( \frac{1}{2} u_p^2 + \frac{1}{2} \log{k_0} \right) \approx \frac{1}{2} B^{\mP} \partial_{\mP} \log{k_0} \,,
\eea
where we have used Eqs.~\eqref{transBb} to relate $b^{\mP} \propto B^{\mP}$ near the launching surface.
Substituting Eq.\eqref{derivativeterm} into Eq.\eqref{LorentzTmn}, we obtain the launching condition:
\bea\label{stagnationpoint}
P^{\prime}\Big|_{\rm LS} + \frac{h}{2}\,\left(\log{k_0} \right)^{\prime} \Big|_{\rm LS} = 0 \,.
\eea
This equation constrains the location of the launching surface, determined by the interplay of gravity, magnetic field structure, and thermal pressure gradients.
In the cold limit, the launching condition reduces to $k_0^{\prime}\big|_{\rm LS} = 0$. This follows directly from the acceleration equation (Eq.\eqref{appacceleration}): to ensure regularity at the launching surface, the numerator $N \propto k_0 k_2 k_0^{\prime}$ must vanish as $u_p \rightarrow 0$.
Using $k_2 = k_0 E^{-2}$ from Eq.\eqref{ELstag} and noting that $k_0 > 0$ within the light cylinder, we conclude that the launching surface must lie at a local extremum of $k_0$.

\section{Kerr light cylinders}\label{Sec:Kerrlightclinders}

Outside a black hole, by numerically solving the condition $k_0 = 0$ for proper values of $\Omega_F$, one obtains two characteristic surfaces \cite{1969ApJ...157..869G, okamoto1992evolution}:
an outer light cylinder, $r = R_{\rm O}(\t)$, located far from the event horizon; an inner light cylinder, $r = R_{\rm I}(\t)$, situated near the horizon, arising from the interplay between the  gravity and magnetic field rotation.

\begin{figure}[ht!]
\hspace*{-3mm}
	\includegraphics[width=7in]{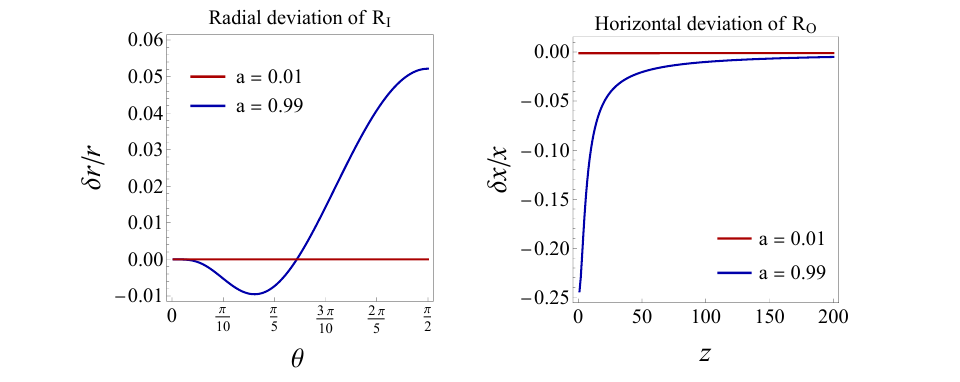}
	\caption{Comparison between the precise and approximate expressions of the light cylinders in Kerr spacetime, assuming $\Omega_F = \Omega_H / 2$, where $\Omega_H = a / (2 r_+)$ is the black hole  angular velocity. The red curves correspond to a low-spin case ($a = 0.01$), and the blue curves correspond to a high-spin case ($a = 0.99$).
\textbf{Left:} radial deviation $\delta r/r \equiv 1 -  R_{\rm I}^{\rm approx}/R_{\rm I}^{\rm precise}$ of the inner light cylinder in polar coordinates. 
\textbf{Right:} horizontal deviation $\delta x/x \equiv 1 -  x_{\rm O}^{\rm approx}/x_{\rm O}^{\rm precise}$ ($x_{\rm O} = R_{\rm O} \sin{\t}$) of the outer light cylinder in cylindrical coordinates.}
	\label{light_cylinder}
\end{figure}

Here, we derive approximate analytical expressions for the locations of the light cylinders in the Kerr spacetime.
By expanding $k_0 = 0$ in terms of $r^{-1}$, we find a simple asymptotic expression:
\bea
R_{\rm O}(\t) \approx \frac{\sqrt{1 - a^2 \Omega_F^2 \sin^2\theta}}{\Omega_F\sin{\t}} + \mathcal{O}\left( \f{1}{r} \right) \,,
\eea
indicating how the outer light cylinder is affected by the black hole spin $a$ and $\Omega_F$.
For the inner light cylinder, by taking the near-horizon expansion for $k_0 = 0$, we have
\bea
R_{\rm I}(\t) \approx r_+ -\frac{
2 r_+ - \Sigma_+ + 4 r_+ \Omega_F (r_+ \Omega_F -a ) \sin^2\theta}
{
2 r_+ \Sigma_+ \Omega_F^2 \sin^2\theta +
2(1-\frac{2r_+^2}{\Sigma_+}) \left( 1 - a \Omega_F \sin^2\theta \right)^2
} + \mathcal{O}\left( r -r_+\right)  \, ,
\eea
where $r_+ =  1 + \sqrt{1-a^2}$ is the event horizon radius, and $\Sigma_+ = r_+^2 + a^2 \cos^2\theta$. In Fig.~\ref{light_cylinder}, we present the relative error between the exact light cylinders and the approximate expression results.
Due to the nearly circular shape of $R_{\rm I}(\t)$, we plot the variation of the relative error in radius with respect to $\t$. From the left panel of Fig.~\ref{light_cylinder}, it can be seen that the approximate expression for the inner light cylinder is highly accurate in low-spin cases, whereas for high spin, it introduces an error of about $2\%$ near the equatorial plane. 

For the outer light cylinder, owing to its cylindrical geometry, we plot the relative error in $x$ as a function of $z$. The approximate expression remains accurate in low-spin cases, but for high-spin black holes, it shows substantial deviations near the equatorial plane, with errors reaching up to $\sim 25\%$.
Despite this, the approximation remains valid in the outflow region, as physically relevant open or collimated magnetic field lines tend to avoid the equator and extend toward higher latitudes.

\section{Jet modeling}
This section provides supplementary material for jet modeling discussed in Sec.~\ref{Sec:jets in Kerr}.

\subsection{The parabolic field}
\label{AppHTMF}
In Boyer-Lindquist coordinates, the poloidal components of the magnetic field corresponding to the stream function Eq.~\eqref{BZstream} are given by
\bea\label{BrBthetaBZ}
B^r = \frac{r^q}{\Sigma}\,, \quad
B^\theta = -\frac{qr^{q-1}(1-\cos{\theta})}{\Sigma \sin{\theta}}\, ,
\eea
where $\Sigma = r^2 + a^2 \cos^2{\theta}$. 
The large-distance behavior of the poloidal field depends on the collimation index. 
For $ q = 0 $, the field lines become radial and the poloidal magnetic field scales as $B_p = \sin^2\theta / x^2$. 
For $0 < q < 2$, the stream function asymptotically satisfies $z = (2\psi)^{1/(q-2)} x^{2/(2-q)}$,
 and the poloidal field behaves as $B_p = 2\psi / x^2$.

\begin{figure}[ht!]
	\centering
	\includegraphics[width=2.8in]{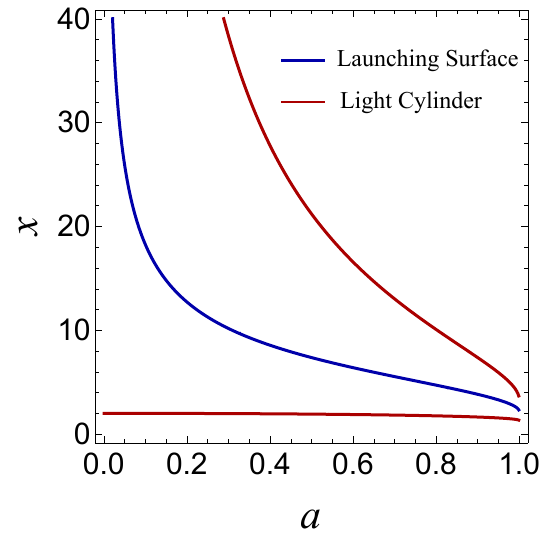}
	\centering
	\caption{
		Dependence of the launching surfaces and light cylinders on black hole spin for the BZ magnetic field configuration. 
		The blue curve indicates the launching surface, defined by $k_0' = 0$; the upper and lower red curves mark the outer and inner light cylinders, 
		defined by $k_0 = 0$. 
		 The magnetic field line with a paraboloidal geometry (i.e., $q = 1$) and anchored at the equator of the event horizon ($\t_+ = \pi/2$).
	}
	\label{BZ_Stag_Light}
\end{figure}

Under the field line angular velocity specified by Eq.\eqref{BZOmega}, the locations of the light cylinders and the launching surface can be determined by solving $k_0 = 0$ and $k_0^{\prime} = 0$, respectively. Fig.~\ref{BZ_Stag_Light} illustrates the dependence of these critical surfaces on the black hole spin, for a representative magnetic field line with paraboloidal geometry (i.e., $q = 1$),  anchored at the equator with $\t_+ = \pi/2$. 
As shown, the frame-dragging effect induced by black hole rotation substantially alters magnetic field rotation. As a result, both the launching surface and the outer light cylinder exhibit pronounced sensitivity to the spin parameter. In contrast, the location of the inner light cylinder—primarily determined by the presence of the black hole itself—remains largely insensitive to variations in spin.

\subsection{The alternative hyperbolic field}\label{TypeIIBP}

As an alternative study to Sec.~\ref{Sec:diskthreadingKerr}, here we consider a hyperbolic field configuration presented in \cite{Gralla:2015vta}, which is a force-free solution. The stream function takes:
\bea\label{BP2psi}
\psi = \psi_0\int \f{u \md u}{\sqrt{\left(1-u^2\right)\left(1-b^2 \Omega_F^2(u)u^4\right)}} \, , \quad u = \frac{\sqrt{(x + b)^2 + z^2} - \sqrt{(x - b)^2 + z^2}}{2 b} \,,
\eea
Typically, $\psi$ is a single-valued function with respect to $u$.
For $u < 1$, the launching point of a given field line is located at $x_0(u) = b u$.
In Boyer-Lindquist coordinates, the poloidal magnetic field components take the form of 
\be
\bag
B^r&= \frac{r \cot\theta}{2\Sigma} \left( \frac{1}{\sqrt{b^2 + r^2 - 2 b r \sin\theta}} +  \frac{1}{\sqrt{b^2 + r^2 + 2 b r \sin\theta}} \right) \f{\md \psi}{\md u}   \,, \\
B^{\t}& = -\frac{1}{2b\Sigma \sin{\t}} \left( \frac{-r + b \sin\theta}{\sqrt{b^2 + r^2 - 2 b r \sin\theta}} +  \frac{r + b \sin\theta}{\sqrt{b^2 + r^2 + 2 b r \sin\theta}} \right)  \f{\md \psi}{\md u}    \,.
\eag
\ee
where $\md \psi / \md u =  u(1-u^2)^{-1/2}(1-b^2 \Omega_F^2(u)u^4)^{-1/2}$.
At large distances, we have the asymptotic relation $u^2z^2 \approx (1-u^2)x^2$, and the field components take
\be
\bag
B^z& \rightarrow \f{u\left(1-u^2\right)}{x^2} \f{\md \psi}{\md u}\,, \quad B^{x}& \rightarrow \f{u^2\sqrt{1-u^2}}{x^2} \f{\md \psi}{\md u}\,.
\eag
\ee
The asymptotic poloidal field magnitude takes
\bea
B_p = \frac{u^2}{\sqrt{(1-b^2\Omega_F^2u^4)}} \frac{1}{x^2} \,.
\eea

To ensure that the field line is physically well-defined, the parameter $u$ must satisfy $0 < u < 1$. 
This constraint follows from Eq.~\eqref{BP2psi}, where $u > 0$ by construction. If $u \geq 1$, the corresponding expression for $z^2$ becomes non-positive. Consequently, the launching radius must satisfy $x_0 < b$ to guarantee that $u < 1$.
Finally, using the expression for the mass flux Eq.~\eqref{etainfity}, we obtain
\bea
\eta = \frac{ \Omega_F^2}{(\gamma_\infty^2 - 1)^{3/2}} \frac{u^2}{\sqrt{(1-b^2\Omega_F^2u^4)}}  \,.
\eea

\begin{figure}[ht!]
	\centering
	\includegraphics[width=6.8in]{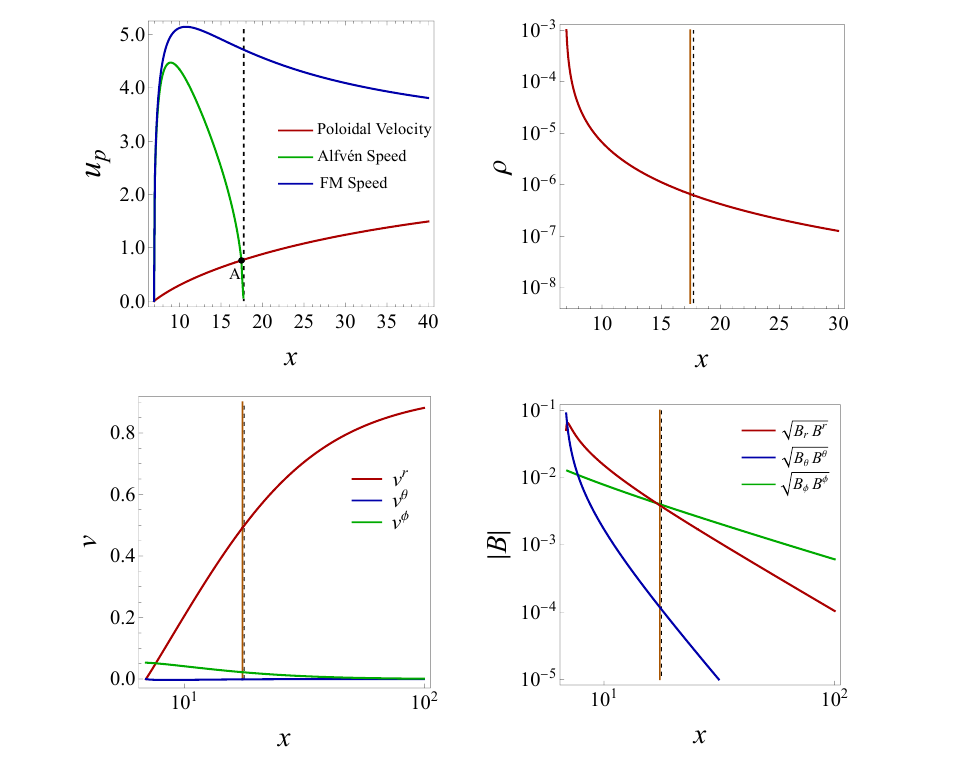}
	\centering
	\caption{
A representative BP flow solution is shown, along a hyperbolic field line given by Eq.~\eqref{BP2psi} with $x_0 = 2.94r_{\rm ISCO}$ and $b = 3r_{\rm ISCO}$. We set the spin $a = 0.9$ and terminal Lorentz factor $\gamma_\infty = 3$. 
	The light cylinder is shown as black dashed lines, and the Alfvén surface as an orange line.
	\textbf{Top-left:}  Poloidal velocity $u_p$ for outflow (red), Alfvén speed (green dashed), and FM speed (blue 
	dashed). The point labeled A indicates the outflow Alfvén points. 
	\textbf{Top-right}: Plasma density.
	\textbf{Bottom-left:}  Components of the three-velocity $v^r$, $v^\theta$, and $v^\phi$. 
	\textbf{Bottom-right:} Magnetic field components $\sqrt{B^r B_r}$, $\sqrt{B^\theta B_\theta}$, and $\sqrt{B^\phi B_\phi}$. 
	}
	\label{BP2_out}
\end{figure}
We illustrate typical features of jet solutions in Fig.~\ref{BP2_out}.
The top-left panel presents the poloidal velocity, the Alfvén speed, and the FM speed of the outflow.
The top-right panel shows the plasma density.
We also plot the three-velocity components and the magnetic field components in the bottom-left and bottom-right panels, respectively.
At large distances, the field has the asymptotic scalings 
\be
\bag
\sqrt{B^r B_r}  &\rightarrow u^2 (1-b^2\Omega_F^2u^4)^{-1/2} /x^2\,, \\
\sqrt{B^{\t} B_{\t}}  &\rightarrow b^2u^5(1-u^2)^{1/2}(1-b^2\Omega_F^2u^4)^{-1/2}/x^4 \,, \\
\sqrt{B^\phi B_\phi}  &\rightarrow u^2 (1-b^2\Omega_F^2u^4)^{-1/2} /x \,, 
\eag
\ee
which also exhibits a strongly toroidal structure. 

\begin{figure}[H]
\hspace*{-1.7cm}
	\includegraphics[width=7.7in]{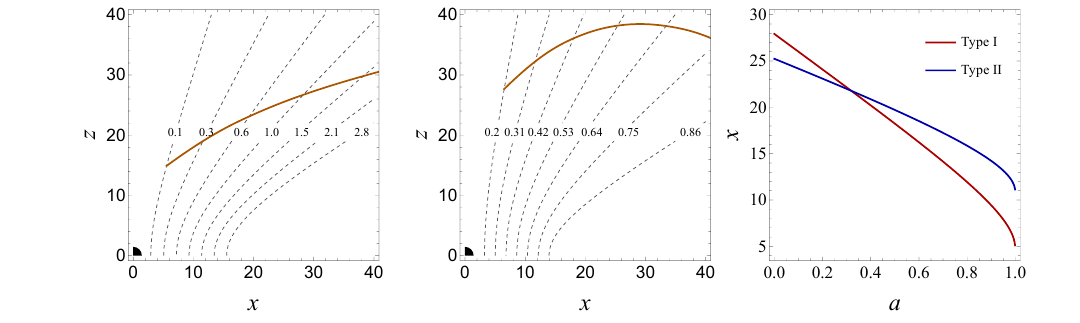}
	\caption{
Comparison of magnetic field structures and light cylinder locations for two types of hyperbolic field configurations.
\textbf{Left:} Type I field with $b = 4r_{\rm ISCO}$.
\textbf{Middle:} Type II field with $b = 7r_{\rm ISCO}$.
Gray dashed lines represent magnetic field lines corresponding to constant values of $\psi$.
The orange curves denote the light cylinders.
\textbf{Right:} Light cylinder radius as a function of black hole spin for both Type I (red) and Type II (blue) configurations. Parameters are set to $b = r_{\rm ISCO} + 1$, $\psi = 1.8$ for Type I, and $b = r_{\rm ISCO} + 5$, $u = 0.8$ for Type II.
	}
	\label{BP12_mag}
\end{figure}

 Fig.~\ref{BP12_mag} compares the poloidal field structures and light cylinder locations generated by two hyperbolic stream functions, given by Eqs.~\eqref{typeIs} (Type I) and \eqref{BP2psi} (Type II). The parameters in each configuration are chosen to produce broadly similar field geometries, facilitating a direct comparison.
The light cylinder in the Type I field exhibits a stronger dependence on black hole spin compared to the Type II case. Nevertheless, for both hyperbolic fields, the spin sensitivity remains weaker than that observed in the parabolic field scenario (see Appendix~\eqref{AppHTMF}). This reduced dependence may be attributed to the anchoring of hyperbolic field lines to the accretion disk, which leads to a weaker coupling between the magnetic field and the black hole.

\clearpage

\subsection{Supplementary figures}
\begin{figure}[H]
	\centering
	\includegraphics[width=6.2in]{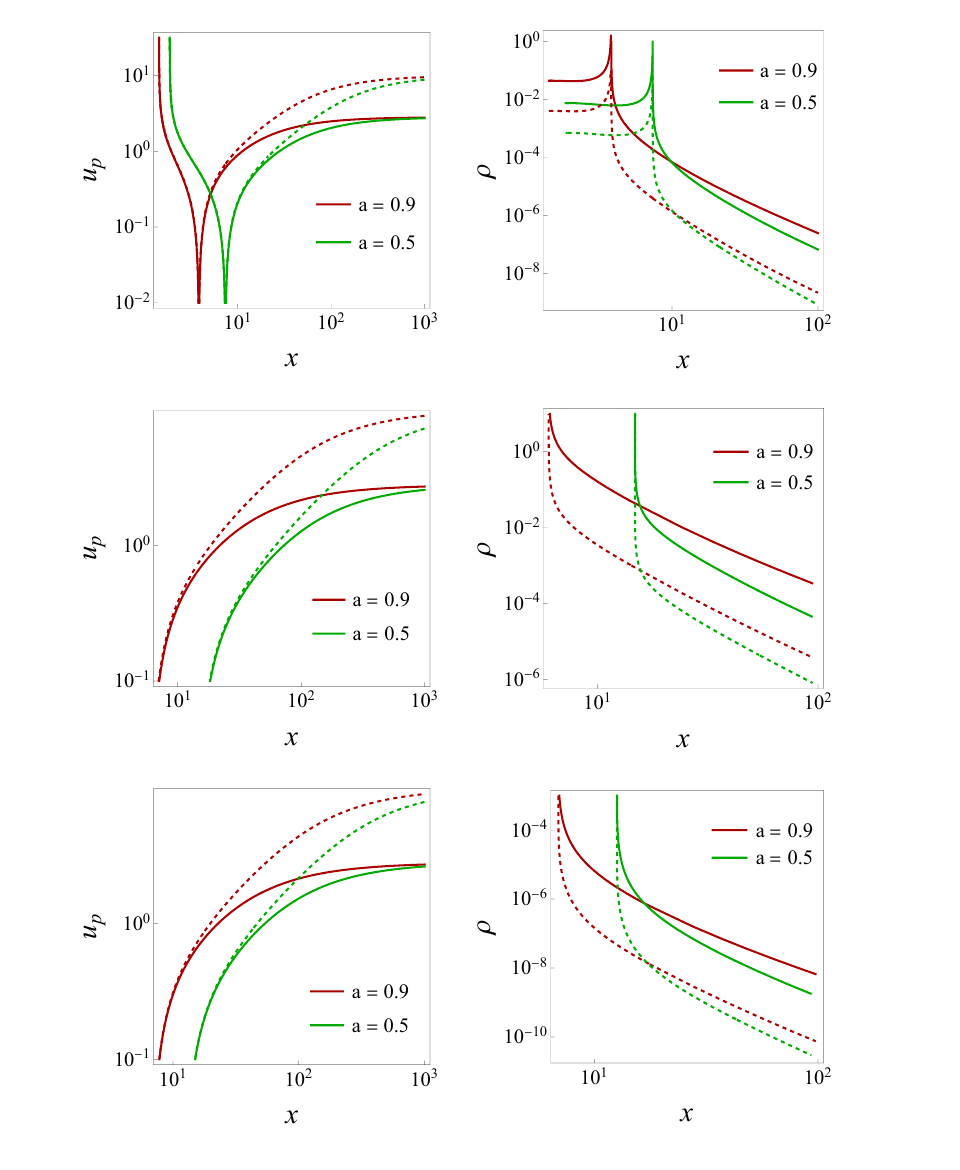}
	\centering
	\caption{
	Poloidal velocity (\textbf{Left}) and plasma density (\textbf{Right}) profiles along magnetic field lines of three different geometries, showing the dependence on black hole spin a and terminal Lorentz factor $\gamma_\infty$.
\textbf{Top row:} Parabolic field with $q = 1$, $\theta_+ = \pi/2$.
\textbf{Middle row:} Type I hyperbolic field anchored at $x_0 = (r_{\rm ISCO} - 1)\sqrt{20}$, with $b = r_{\rm ISCO} - 1$.
\textbf{Bottom row:} Type II hyperbolic field anchored at $x_0 = 2.94r_{\rm ISCO}$, with $b = 3r_{\rm ISCO}$.
In all panels, solid lines correspond to $\gamma_\infty = 3$, dashed lines to $\gamma_\infty = 10$.
Red curves represent spin $a = 0.9$; green curves represent $a = 0.5$.
	}
	\label{sup_uprho}
\end{figure}

As a complement to the main text, we present the poloidal velocity and density profiles for different magnetic field configurations in Fig.~\ref{sup_uprho}, in order to examine how the jet solutions depend on the black hole spin and the terminal Lorentz factor.

From top to bottom, the rows correspond to the parabolic, Type I hyperbolic, and Type II hyperbolic magnetic field configurations, respectively. In the parabolic case, the primary distinctions among the cases stem from the differing locations of the launching surfaces, which are governed by variations in black hole spin. 
Additionally, differences in the terminal Lorentz factor $\gamma_\infty$ lead to distinct asymptotic behaviors in the outflow poloidal velocities. In contrast, the inflow solutions exhibit minimal sensitivity to $\gamma_\infty$, rendering the differences between the two Lorentz factors barely discernible. The right panels show the corresponding density profiles. The dominant variations arise from shifts in the launching surfaces and changes in the overall density normalization. For the two hyperbolic configurations, both the velocity and density profiles share nearly identical shapes, with the primary difference in density being a uniform scaling factor.

\bibliographystyle{utphys}
\bibliography{jet_draft}

\end{document}